\documentclass[pra,aps,twocolumn]{revtex4}

\usepackage{amssymb}
\usepackage{amsmath}
\usepackage{epsfig}

\begin{document}
\title{Classification of GHZ-type, W-type and GHZ-W-type multiqubit entanglements}
\author{Lin Chen}
\author{Yi-Xin Chen}
\affiliation{Zhejiang Insitute of Modern Physics, Zhejiang
University, Hangzhou 310027, China}

\begin{abstract}
We propose the concept of SLOCC-equivalent basis (SEB) in the
multiqubit space. In particular, two special SEBs, the GHZ-type
and the W-type basis are introduced. They can make up a more
general family of multiqubit states, the GHZ-W-type states, which
is a useful kind of entanglement for quantum teleporatation and
error correction. We completely characterize the property of this
type of states, and mainly classify the GHZ-type states and the
W-type states in a regular way, which is related to the
enumerative combinatorics. Many concrete examples are given to
exhibit how our method is used for the classification of these
entangled states.
\end{abstract}
\maketitle

\section{Introduction}
The recent development of quantum information theory (QIT) shows
the importance of entanglement. As a kind of quantum nonlocal
correlation, entanglement has been used to many new tasks which
are unpractical or even impossible in the classical scenarios
\cite{Bennett1,Chuang}. The study of entanglement is thus greatly
inspired by these novel phenomena, and much progress has been made
in quantifying and classifying entanglement in both bipartite and
multipartite system \cite{Bennett2,Linden,Vidal}. In spite of lots
of effort, a better understanding of multiple entanglement is
still required, which contains a much richer configuration than
the bipartite setting \cite{Eisert}.

In recent years, a subject of most interest in QIT is to
characterize the entanglement properties of multiqubit states
\cite{Dur2,Wootters,Stockton}. The main motivation is that the
mutiqubit entangled states play the role of quantum correlations
in a quantum computer, since any unitary operation could be
decomposed into a series of two-level unitary operations
\cite{Chuang,Mottonen}, each of which act on one spin-$1/2$
particle. On the other hand, the latest experimental progress
shows the multiqubit entanglement is also a kind of practically
physical resource, such as the decoherence-free quantum
information processing \cite{Bourennane}. In particular, the
characterization of two important kinds of 3-qubit states, the
Greenberger-Horne-Zeilinger (GHZ) state,
$\left|GHZ\right\rangle=(\left|000\right\rangle+\left|111\right\rangle)/\sqrt{2}$,
and W state,
$\left|W\right\rangle=(\left|001\right\rangle+\left|010\right\rangle+\left|100\right\rangle)/\sqrt{3}$
has been realized by using quantum state tomography
\cite{Dur1,Resch,Mikami}. By virtue of these facts, one can expect
to generally construct and manipulate macroscopic quantum states.
In this sense, the multiqubit states are considered as the most
fundamental quantum resource.

The case becomes complicated when trying to classify the
multiqubit entanglement under a specific constraint, e.g., local
operations and classical communications (LOCC), because of the
exponentially increasing number of parameters \cite{Linden}. Under
the LOCC criterion, two pure states $\left|\psi\right\rangle$ and
$\left|\phi\right\rangle$ are interconvertible with certainty if
and only if (iff) they are related by local unitary operations
\cite{Vidal}. The meaning of classification is that the states
belonging to the same family can be used to finish the same
quantum-information task, possibly with a different probability.
Although the LOCC criterion can explicitly judge the equivalence
of two states, it is exceedingly inconvenient when classifying
multiple entanglement \cite{Linden,Acin1}. For simplicity, a more
useful criterion requires that two states are regarded as
equivalent by LOCC in a stochastic manner (SLOCC), i.e., related
by an invertible local operator (ILO) only with a nonvanishing
probability \cite{Dur1,Bennett3,Verstraete1}. As far as the SLOCC
criterion is concerned, GHZ state and W state are the only two
classes of states of the genuine 3-qubit system \cite{Dur1}. These
two types of states (including the generalized GHZ and W state of
multipartite system) have been applied to many quantum-information
process, such as the quantum secret sharing \cite{Hillery},
quantum key distribution \cite{Kempe} and quantum teleportation
\cite{Yang,Rigolin,Deng,Ye}. Noticeably, the scheme of many-party
controlled teleportation in \cite{Yang} employed a ``quasi" GHZ
state like this,
\begin{eqnarray}
\prod^m_{i=1}(\left|00\right\rangle_{i^{\prime}i^{\prime\prime}}+\left|11\right\rangle_{i^{\prime}i^{\prime\prime}})
\otimes(\left|00\right\rangle_{ac}+\left|11\right\rangle_{ac})+\nonumber\\
\prod^m_{i=1}(\left|00\right\rangle_{i^{\prime}i^{\prime\prime}}-\left|11\right\rangle_{i^{\prime}i^{\prime\prime}})
\otimes(\left|00\right\rangle_{ac}-\left|11\right\rangle_{ac}),
\end{eqnarray}
which can be viewed as a combination of the EPR pair of different
bipartite system. This type of state more frequently appears in
the error correction, e.g., the four-qubit code \cite{Vaidman},
which corrects one erasure by the encoding
\begin{eqnarray}
\alpha\left|0\right\rangle+\beta\left|1\right\rangle\rightarrow\frac{\alpha}2
(\left|00\right\rangle+\left|11\right\rangle)(\left|00\right\rangle+\left|11\right\rangle)+\nonumber\\
\frac{\beta}2(\left|00\right\rangle-\left|11\right\rangle)(\left|00\right\rangle-\left|11\right\rangle).
\end{eqnarray}
Due to these facts, the states composed of the GHZ states (and W
states) can be regarded as an important kind of quantum resource.
On the other hand, such a state has a simple configuration
compared to other multiqubit states, so it is likely to give a
complete investigation of it.

In the following we will characterize this kind of entanglement.
Different from the existing results such as the 4-qubit
entanglement \cite{Verstraete2}, our work is devoted to the
classification of a general family of pure multiqubit states under
SLOCC criterion (we shall use $\sim$ to denote the SLOCC
equivalence), by constructing the so-called SLOCC-equivalent
basis. Given the number of parties in the system, this family can
be seen as a subset of corresponding multiqubit entangled classes.
In particular, the family of states will be regularly classified
into infinitely many subclasses, each of which contains certain
parameters. It is shown that the general classification of the
family requires the technique of enumerative combinatorics
\cite{Stanley}. Furthermore, we give out the relative ILO's making
one class into another, and one can efficiently remove the
parameters to obtain a brief expression of every class. So our
classification is a more explicit scheme than that in
\cite{Verstraete2}. This helps completely classify the general
multiqubit states. Besides, it is easy to convert our results into
the explicit classification under LOCC.

In section II we recall the range criterion \cite{Chen}, based on
which the SLOCC equivalent basis (SEB) is proposed. We explain the
meaning of SEB and particularly construct two general kinds of
multiqubit SEBs, the GHZ-type basis and W-type basis, which is two
special cases of the rank-2 basis. We prove that the GHZ state is
the only type consisting of the GHZ-type basis in each two-qubit
subspace, while the W state is the only one consisting of the
W-type basis in each two-qubit subspace. Subsequently, we exhibit
a new family of pure multiqubit states, i.e., the GHZ-W-type
state, which is composed of the GHZ-type and W-type basis. We
characterize its entanglement property. First we present the
condition of full entanglement, since our target should not be
separable. In particular, the correspondence relation greatly
simplifies the traditional procedure of entanglement detection.
Besides, we also give a practical criterion to distinguish whether
a general multiqubit state with unknown coefficient could be fully
entangled. Second, we explore the simplest form of the GHZ-W-type
state by using of the GHZ-criterion and W-criterion newly
established. Moreover, we calculate the ways of partitions into
which an N-qubit GHZ-W-type state can be divided by the theory of
combinatorics. By virtue of these techniques, we first classify
the GHZ-type state under the SLOCC criterion in section III, where
some useful notions are proposed, such as the relative ILO and
column. These tricks can be applied to catalog the W-type state,
while the general GHZ-W-type state can be viewed as the mixture of
the GHZ-type and the W-type state, as shown in section IV. We
describe how to explicitly write out the inequivalent classes of
states by many typical examples. The conclusion is given in
section V.

\section{GHZ-type basis, W-type basis and GHZ-W-type multiqubit entanglement}

Several leading theories of entangled bases have been established
in QIT. For example, there is an intimate correlation between the
unextendible product bases (UPBs) and the bound entanglement
\cite{Bennett4}, while the mutually unbiased bases (MUBs)
\cite{Romero} has also been particularly investigated since they
can be used for the the discrete Wigner function \cite{Gibbons}
and mean king problem \cite{Hayashi}. In the present work, we
apply the entangled bases to the classification of true multiqubit
entangled states under SLOCC, by virtue of the range criterion
\cite {Chen}. Let $\rho$ be an operation on space ${\cal H}$, then
the range of $\rho$ is defined by
$R(\rho)=\rho\left|\Phi\right\rangle$, for some
$\left|\Phi\right\rangle\in {\cal H}$ \cite{Horn}. For generality,
a density operator $\rho$ can be written as
$\rho=\sum_{i=0}^{n-1}{p_i\left|\psi_i\right\rangle\left\langle\psi_i\right|}$,
where $\{\left|\psi_i\right\rangle,i=0,1,...,n-1\}$ are a set of
linearly independent vectors. It is thus easy to see that
\begin{equation}
R(\rho)=\sum_{i=0}^{n-1}{p^\prime_i\left|\psi_i\right\rangle},
p^\prime_i=p_i\left\langle\psi_i|\Phi\right\rangle.
\end{equation}
In this case, $R(\rho)$ can be regarded as a vector space spanned
by a set of basis $\{\left|\psi_i\right\rangle,i=0,1,...,n-1\}$,
which is a set of entangled bases. Define the reduced density
operator $\rho^{A_{ik+1},A_{ik+2},\cdots,A_{iN}}_\Psi\equiv$
tr$_{A_{i1},A_{i2},\cdots,A_{ik}}(\rho_{_{\scriptstyle \Psi}}),
i_1,i_2,\cdots,i_k\in\{1,2,\cdots,N\}, k\leq{N-1}$. It has been
shown that the range of the reduced density operator of a
multipartite pure state
$\left|\Psi\right\rangle_{{A_1}{A_2}\cdots{A_N}}$ essentially
determines the sort of this state,

\textit{Range Criterion}. Consider two multiple states
$\left|\Psi\right\rangle_{{A_1}{A_2}\cdots{A_N}}$ and
$\left|\Phi\right\rangle_{{A_1}{A_2}\cdots{A_N}}$ with the
identical local rank of each party. Then there exist certain ILO's
$V_i, i=1,...,N$ making
$\left|\Psi\right\rangle_{{A_1}{A_2}\cdots{A_N}}=
{V_1}\otimes{V_2}\otimes\cdots\otimes{V_N}\left|\Phi\right\rangle_{{A_1}{A_2}\cdots{A_N}}$,
iff $
R(\rho^{A_{i1},A_{i2},\cdots,A_{i_{N-1}}}_\Psi)={V_{i1}}\otimes{V_{i2}}\otimes\cdots\otimes{V_{i_{N-1}}}
R(\rho^{A_{i1},A_{i2},\cdots,A_{i_{N-1}}}_\Phi),i_1,i_2,\cdots,i_{N-1}\in\{1,2,\cdots,N\}$.
\hspace*{\fill}$\blacksquare$

The range criterion has been used for the classification of
several families of true multipartite states, such as the
$2\times3\times N$ and $2\times4\times4$ entanglement containing
finite and infinite kinds of states respectively \cite{Chen}. A
prominent feature of these existing results is that most of them
have distinct local ranks, which greatly helps distinguish the
correspondence relation of parties since the ILO's cannot alter
the local rank \cite{Dur1}. For example, if two $2\times3\times4$
states $\left|\Psi_0\right\rangle_{ABC}$ and
$\left|\Psi_1\right\rangle_{A^{\prime}B^{\prime}C^{\prime}}$ are
equivalent under SLOCC, then the correspondence relation must be
$A-A^{\prime}$, $B-B^{\prime}$ and $C-C^{\prime}$. In other word,
we only need to consider the structure of the ranges, e.g., the
number of product states therein. While in the case of multiqubit
system, all parties have the same local rank two. It thus becomes
difficult to distinguish the correspondence relation when
classifying this crucial kind of entanglement. However, we will
show that the entangled bases can help explore the structure of
multiqubit states.

Suppose $\rho_1$ and $\rho_2$ are two reduced density operators of
states $\left|\Psi_1\right\rangle$ and $\left|\Psi_2\right\rangle$
respectively.  If they have the same local ranks, then the
equivalence of them under SLOCC implies $R(\rho_1)\sim R(\rho_2)$
due to the range criterion. We thus can say that $R(\rho_1)$ and
$R(\rho_2)$ are equivalent spaces under SLOCC, i.e., they are
spanned by the SLOCC-equivalent basis (SEB). Concretely, let
$S_0=\{\left|\psi_i\right\rangle_{{A_1}{A_2}\cdots{A_N}},i=0,1,...,n-1\}$
be a set of bases, then its SEB must have the form
$S_1=\{\left|\phi_i\right\rangle_{{A_1}{A_2}\cdots{A_N}}={V_1}\otimes{V_2}\otimes\cdots\otimes{V_N}
\sum_{j=0}^{n-1}a_{ij}\left|\psi_j\right\rangle_{{A_1}{A_2}\cdots{A_N}},i=0,1,...,n-1\}$,
the coefficient matrix $A^{n\times n}=[a_{ij}]$ is nonsingular and
$V_i, i=1,...,N$ are any ILO's. Conversely, it is easy to see that
$S_1$ is also a set of SEB of $S_0$, so $S_0\sim S_1$. Since the
SLOCC-equivalent states must have the same SEB, we can catalog the
multipartite entanglement by finding out all different SEBs. For
example, \cite{Sampera} has proved that a two-dimensional subspace
of $\mathcal{C}^2\otimes\mathcal{C}^2$ always contains either only
one or only two product vectors. This has been used for the
classification of 3-qubit states \cite{Dur1}, where they indeed
offered the explicit form of two simplest SEBs,
$R_{20}\equiv\{\left|01\right\rangle+\left|10\right\rangle,\left|00\right\rangle\},
R_{21}\equiv\{\left|00\right\rangle,\left|11\right\rangle\}$,
containing one and two product states respectively. Due to the
above argument, the set of GHZ class can be expressed as
\begin{eqnarray}
\left|\Psi_{GHZ}\right\rangle_{ABC}&=&
V_A\otimes V_B(a_{00}\left|00\right\rangle+a_{01}\left|11\right\rangle)_{AB}\left|0\right\rangle_C\nonumber\\
&+&V_A\otimes
V_B(a_{10}\left|00\right\rangle+a_{11}\left|11\right\rangle)_{AB}\left|1\right\rangle_C,
\end{eqnarray}
where $a_{00}a_{11}-a_{01}a_{10}\neq0$, $V_A$ and $V_B$ are
nonsingular. Clearly, any state SLOCC-equivalent to the GHZ state
has the above form. In the same vein one can write out the
expression of the W class $\left|\Psi_{W}\right\rangle_{ABC}$.

The SEBs $R_{20}$ and $R_{21}$ are two cases of the rank-2 basis
of the multiqubit space, i.e., either of them only contains two
independent vectors (they are also the only two SEBs in the true
2-qubit space). It is natural to refer to the rank-k basis as
those including k independent vectors. In general, any state can
be seen as a synthesis consisting of the rank-k basis in each
subspace of several parties, e.g., $\left|GHZ\right\rangle_{ABC}$
is composed of $R_{21}$ spanning $AB$, $AC$ and $BC$ spaces.
Generalize the SEBs $R_{20}$ and $R_{21}$ to the multipartite
case, i.e., define the W-type basis
$R_{N0}\equiv\{\left|0\rangle\right\rangle_N,\left|W\right\rangle_N\}$
and the GHZ-type basis
$R_{N1}\equiv\{\left|0\rangle\right\rangle_N,\left|1\rangle\right\rangle_N\}$,
where
\begin{eqnarray}
\left|0\rangle\right\rangle_N&\equiv&|\overbrace{0,0,...,0}^{N}\rangle,\
\
\left|1\rangle\right\rangle_N\equiv|\overbrace{1,1,...,1}^{N}\rangle,\nonumber\\
\left|W\right\rangle_N&\equiv&|\overbrace{1,0,...,0}^{N}\rangle
+\left|0,1,...,0\right\rangle+\cdots+\left|0,0,...,1\right\rangle,\nonumber\\
N\geq2.
\end{eqnarray}
Evidently, $\left|W\right\rangle_N$ is the N-partite W state and
anther well-known N-qubit state has the form
$\left|GHZ\right\rangle_N\equiv\left|0\rangle\right\rangle_N+\left|1\rangle\right\rangle_N$.
Because the space spanned by $R_{N0}$ contains a unique fully
product state and $R_{N1}$ has two, they are inequivalent under
SLOCC. The states containing the two SEBs can be respectively
written as
\begin{eqnarray}
\left|\Omega_{N0}\right\rangle&\equiv&\left|W\right\rangle_N\left|\psi_0\right\rangle
+\left|0\rangle\right\rangle_N\left|\phi_0\right\rangle,\\
\left|\Omega_{N1}\right\rangle&\equiv&\left|0\rangle\right\rangle_N\left|\psi_1\right\rangle
+\left|1\rangle\right\rangle_N\left|\phi_1\right\rangle,
\end{eqnarray}
where each pair of
$\left|\psi_i\right\rangle,\left|\phi_i\right\rangle,i=0,1$, which
are linearly independent, denote the states belonging to the
Hilbert space of extra parties. As both of $R_{N0}$ and $R_{N1}$
are symmetric under particle exchange, we always have
\begin{eqnarray}
\left|\Omega_{N0}\right\rangle&=&
[(\left|01\right\rangle+\left|10\right\rangle)_{AB}\left|0\rangle\right\rangle_{N-2}+
\left|00\right\rangle_{AB}\left|W\right\rangle_{N-2}]\left|\psi_0\right\rangle\nonumber\\
&&+\left|00\right\rangle_{AB}\left|0\rangle\right\rangle_{N-2}\left|\phi_0\right\rangle\nonumber\\
&=&
(\left|01\right\rangle+\left|10\right\rangle)_{AB}\left|\psi^{\prime}_0\right\rangle
+\left|00\right\rangle_{AB}\left|\phi^{\prime}_0\right\rangle\nonumber\\
&\sim&\left|\Omega_{20}\right\rangle,
\end{eqnarray}
where $A,B$ are randomly chosen from the first $N$ parties.
Similarly, $\left|\Omega_{N1}\right\rangle$ is a special case of
$\left|\Omega_{21}\right\rangle$. Notice that both of the states
$\left|W\right\rangle_N$ and $\left|GHZ\right\rangle_N$ consist of
the rank-2 basis ($R_{20}$ and $R_{21}$ respectively) in each
two-qubit subspace. We consider the converse of this observation.
That is, do there exist other multiqubit states composed of
$R_{20}$ or $R_{21}$ in each two-qubit subspace, besides the
$\left|GHZ\right\rangle_N$ and $\left|W\right\rangle_N$? The
answer is phrased as follows.

\textit{Theorem 1}. Suppose the multiqubit state
$\left|\Psi\right\rangle_{{A_1}{A_2}\cdots{A_N}}$ consists of
$R_{20}$ or $R_{21}$ in each two-qubit subspace, then
$\left|\Psi\right\rangle_{{A_1}{A_2}\cdots{A_N}}$ is equivalent to
either $\left|GHZ\right\rangle_N$ or $\left|W\right\rangle_N$
under SLOCC.

\textit{Proof}. Two existing ILO's \cite{Chen} are useful
henceforth. They are defined by
$O^A_1(\left|\phi\right\rangle,\alpha):
\left|\phi\right\rangle_A\rightarrow\alpha\left|\phi\right\rangle_A$
and $O^A_2(\left|\phi\right\rangle,\left|\psi\right\rangle):
\left|\phi\right\rangle_A\rightarrow\left|\phi\right\rangle_A+\left|\psi\right\rangle_A$,
respectively. Besides, we will omit the specific form of ILO's in
the deduction if unnecessary. First consider the separable state,
which is the direct product of at least two states of different
systems, i.e., $\left|\Psi\right\rangle_{{A_1}{A_2}\cdots{A_N}}=
\bigotimes\prod_i\left|\Psi_i\right\rangle$, with every
$\left|\Psi_i\right\rangle$ truly entangled. Choose the first two
states and we can always write
$\left|\Psi_0\right\rangle_{AB}\left|\Psi_1\right\rangle_{CD}
=(\left|0\right\rangle\left|\psi_0\right\rangle+\left|1\right\rangle\left|\psi_1\right\rangle)_{AB}
(\left|0\right\rangle\left|\phi_0\right\rangle+\left|1\right\rangle\left|\phi_1\right\rangle)_{CD}$,
where $B$ (and $D$) can be composite system. If
$\left|\psi_0\right\rangle$ and $\left|\psi_1\right\rangle$,
$\left|\phi_0\right\rangle$ and $\left|\phi_1\right\rangle$ are
linearly independent respectively, then $R(\rho^{AC}_{ABCD})$ is
spanned by the rank-4 SEB. For the case of linear dependence of
one pair, e.g., $\left|\psi_0\right\rangle$ and
$\left|\psi_1\right\rangle$, we have $R(\rho^{AC}_{ABCD})\sim
a\left|00\right\rangle+b\left|01\right\rangle$. Hence this
separable state is not fully spanned by $R_{20}$ or $R_{21}$.
Next, let us move to
the case of true entanglement, which has two subcases.\\
(i) The states containing $R_{20}$, which must be in the form
$\left|\Omega_{20}\right\rangle=(\left|01\right\rangle+\left|10\right\rangle)\left|\psi_0\right\rangle
+\left|00\right\rangle\left|\phi_0\right\rangle$. We show that it
must also be $
\left|\Omega_{20}\right\rangle\sim\left|W\right\rangle_N\left|\psi^{\prime}_0\right\rangle
+\left|0\rangle\right\rangle_N\left|\phi^{\prime}_0\right\rangle$
by induction. Evidently, $\left|\Omega_{20}\right\rangle$ is the
case of $N=2$ and we can write
\begin{eqnarray}
\left|\Omega_{N0}\right\rangle&=&\left|W\right\rangle_N
(\left|0\right\rangle_B\left|\psi_{00}\right\rangle+\left|1\right\rangle_B\left|\psi_{01}\right\rangle)\nonumber\\
&&+\left|0\rangle\right\rangle_N
(\left|0\right\rangle_B\left|\psi_{10}\right\rangle+\left|1\right\rangle_B\left|\psi_{11}\right\rangle)\nonumber\\
&=&\left|00\right\rangle_{AB}
(\left|W\right\rangle_{N-1}\left|\psi_{00}\right\rangle+
\left|0\rangle\right\rangle_{N-1}\left|\psi_{10}\right\rangle)\nonumber\\
&&+\left|01\right\rangle_{AB}(\left|W\right\rangle_{N-1}\left|\psi_{01}\right\rangle
+\left|0\rangle\right\rangle_{N-1}\left|\psi_{11}\right\rangle)\nonumber\\
&&+\left|10\right\rangle_{AB}\left|0\rangle\right\rangle_{N-1}\left|\psi_{00}\right\rangle
+\left|11\right\rangle_{AB}\left|0\rangle\right\rangle_{N-1}\left|\psi_{01}\right\rangle,\nonumber\\
\end{eqnarray}
where $A$ is one of the first $N$ particles. To keep $AB$ space is
spanned by $R_{20}$ or $R_{21}$, there must be
\begin{eqnarray}
\alpha(\left|W\right\rangle_{N-1}\left|\psi_{00}\right\rangle+
\left|0\rangle\right\rangle_{N-1}\left|\psi_{10}\right\rangle)\nonumber\\
+\beta(\left|W\right\rangle_{N-1}\left|\psi_{01}\right\rangle
+\left|0\rangle\right\rangle_{N-1}\left|\psi_{11}\right\rangle)\nonumber\\
+\gamma\left|0\rangle\right\rangle_{N-1}\left|\psi_{00}\right\rangle
+\delta\left|0\rangle\right\rangle_{N-1}\left|\psi_{01}\right\rangle&=&0,
\end{eqnarray}
or more explicitly
\begin{equation}
\alpha\left|\psi_{00}\right\rangle+\beta\left|\psi_{01}\right\rangle=0,
\end{equation}
and
\begin{equation}
\alpha\left|\psi_{10}\right\rangle+\beta\left|\psi_{11}\right\rangle+
\gamma\left|\psi_{00}\right\rangle+\delta\left|\psi_{01}\right\rangle=0.
\end{equation}
Observe the two equations. If $\left|\psi_{00}\right\rangle$ and
$\left|\psi_{01}\right\rangle$ are linearly independent, then it
has to be $\alpha=\beta=\gamma=\delta=0$, which means $AB$ space
is spanned by rank-4 basis. It follows that
$\left|\psi_{00}\right\rangle$ and $\left|\psi_{01}\right\rangle$
must be linearly dependent. By some ILO's $O^B_1$ and $O^B_2$, we
obtain
\begin{equation}
\left|\Omega_{N0}\right\rangle\sim\left|W\right\rangle_N
\left|0\right\rangle_B\left|\psi^\prime_{00}\right\rangle+\left|0\rangle\right\rangle_N
(\left|0\right\rangle_B\left|\psi^\prime_{10}\right\rangle+\left|1\right\rangle_B\left|\psi^\prime_{11}\right\rangle).
\end{equation}
Analyzing this equation in the same vein gives rise to
$\left|\psi^\prime_{11}\right\rangle=k\left|\psi^\prime_{00}\right\rangle$,
where $k$ is some universal factor. Afterwards, We perform an
operation $O^B_1(\left|1\right\rangle,1/k)$ and have
\begin{eqnarray}
\left|\Omega_{N0}\right\rangle&\sim&\left|W\right\rangle_N
\left|0\right\rangle_B\left|\psi^\prime_{00}\right\rangle+\left|0\rangle\right\rangle_N
(\left|0\right\rangle_B\left|\psi^\prime_{10}\right\rangle
+\left|1\right\rangle_B\left|\psi^\prime_{00}\right\rangle)\nonumber\\
&=&\left|W\right\rangle_{N+1}\left|\psi^\prime_{00}\right\rangle
+\left|0\rangle\right\rangle_{N+1}\left|\psi^\prime_{10}\right\rangle=\left|\Omega_{N+1,0}\right\rangle.
\end{eqnarray}
Therefore, the states containing $R_{N0}$ must also be in the form
of $\left|\Omega_{N+1,0}\right\rangle$. Continue the above
procedure and at last we arrive at the case in which
$\left|\psi^\prime_{00}\right\rangle$ is local. By some ILO we
obtain
$\left|\Omega_{20}\right\rangle\sim\left|W\right\rangle_{N^\prime-1}\left|0\right\rangle
+\left|0\rangle\right\rangle_{N^\prime-1}\left|1\right\rangle=\left|W\right\rangle_{N^\prime}$.
This result asserts that the N-partite W state
is the unique one composed of $R_{20}$ in each two-qubit space of N particle system.\\
(ii) The states containing $R_{21}$, which must be in the form
$\left|\Omega_{21}\right\rangle=\left|00\right\rangle\left|\psi_0\right\rangle
+\left|11\right\rangle\left|\psi_1\right\rangle$. One can do it
following the technique in (i). Write out
\begin{eqnarray}
\left|\Omega_{N1}\right\rangle&=&\left|0\rangle\right\rangle_N
(\left|0\right\rangle_B\left|\psi_{00}\right\rangle+\left|1\right\rangle_B\left|\psi_{01}\right\rangle)\nonumber\\
&&+\left|1\rangle\right\rangle_N
(\left|0\right\rangle_B\left|\psi_{10}\right\rangle+\left|1\right\rangle_B\left|\psi_{11}\right\rangle)\nonumber\\
&=&\left|00\right\rangle_{AB}\left|0\rangle\right\rangle_{N-1}\left|\psi_{00}\right\rangle
+\left|01\right\rangle_{AB}\left|0\rangle\right\rangle_{N-1}\left|\psi_{01}\right\rangle\nonumber\\
&&+\left|10\right\rangle_{AB}\left|1\rangle\right\rangle_{N-1}\left|\psi_{10}\right\rangle
+\left|11\right\rangle_{AB}\left|1\rangle\right\rangle_{N-1}\left|\psi_{11}\right\rangle,\nonumber\\
\end{eqnarray}
and thus
\begin{eqnarray}
\alpha\left|\psi_{00}\right\rangle+\beta\left|\psi_{01}\right\rangle&=&0,\\
\gamma\left|\psi_{10}\right\rangle+\delta\left|\psi_{11}\right\rangle&=&0.
\end{eqnarray}
By the reason similar to that in (i),
$\left|\psi_{01}\right\rangle$ disappears and thus
$\left|\psi_{10}\right\rangle,\left|\psi_{11}\right\rangle$ are
linearly dependent. After an operation $O^B_2$ one finds that
\begin{equation}
\left|\Omega_{21}\right\rangle
\sim\left|0\rangle\right\rangle_{N+1}\left|\psi^\prime_{00}\right\rangle
+\left|1\rangle\right\rangle_{N+1}\left|\psi^\prime_{10}\right\rangle.
\end{equation}
Again we iterate this procedure and finally obtain
$\left|\Omega_{21}\right\rangle\sim\left|GHZ\right\rangle_{N^\prime}$,
so the N-partite GHZ state is the unique one composed of $R_{21}$
in each two-qubit subspace of N particle system. This completes
the proof. \hspace*{\fill}$\blacksquare$

Theorem 1 says that for the N-partite system,
$\left|GHZ\right\rangle_N$ and $\left|W\right\rangle_N$ are the
only two types consisting of rank-2 basis in each two-qubit
subspace under SLOCC criterion. Conversely, this also implies that
a multiqubit state containing both $R_{20}$ and $R_{21}$ must
contain higher rank basis. For example, consider the following
state,
\begin{equation}
\left|\Psi\right\rangle=\left|000\right\rangle_{ABC}(\left|01\right\rangle+\left|10\right\rangle)_{DE}
+\left|111\right\rangle_{ABC}\left|00\right\rangle_{DE}.
\end{equation}
Here, $R(\rho^{ABC}_{\Psi})$ and $R(\rho^{DE}_{\Psi})$ are spanned
by $R_{31}$ and $R_{20}$ respectively. One can briefly check that
$R(\rho^{CD}_{\Psi})$ is spanned by the rank-3 basis, so
$\left|\Psi\right\rangle$ equals neither the GHZ state nor the W
state under SLOCC. Nonetheless, it remains a simple type of
multiqubit entanglement for it consists of the rank-2 basis. As
mentioned above, a general multiqubit state can be seen as the
combination of the rank-k basis in each subspace of several
parties. Intuitively, the states consisting of the rank-2 basis
have a simpler structure compared to those composed of the rank-3
basis, etc, and it is always a wise decision to first treat the
easier cases of a difficult problem. As the GHZ-type and W-type
basis are two special kinds of rank-2 SEBs \cite{notation1}, we
will particularly characterize a general family of multiqubit
states consisting of these two types of SEBs. Differing from other
existing results such as the 3-qubit and 4-qubit system, our
classification describes a subclass belonging to any multiqubit
entanglement, so it helps get insight into the structure of a
general multiqubit state. Let this family of states be
$\left|\Omega_{GHZ-W}\right\rangle_{k,n-k}, 0\leq k\leq n$,
which is defined in the following form\\
\begin{eqnarray}
&&\left|0\rangle\right\rangle_{N_0}\left|0\rangle\right\rangle_{N_1}
\cdots\left|0\rangle\right\rangle_{N_{k-2}}\left|0\rangle\right\rangle_{N_{k-1}}\nonumber\\
&&(a_{0}\left|0\rangle\right\rangle_{N_k}\left|0\rangle\right\rangle_{N_{k+1}}
\cdots\left|0\rangle\right\rangle_{N_{n-2}}\left|0\rangle\right\rangle_{N_{n-1}}\nonumber\\
&&+a_{1}\left|0\rangle\right\rangle_{N_k}\left|0\rangle\right\rangle_{N_{k+1}}
\cdots\left|0\rangle\right\rangle_{N_{n-2}}\left|W\right\rangle_{N_{n-1}}\nonumber\\
&&+a_{2}\left|0\rangle\right\rangle_{N_k}\left|0\rangle\right\rangle_{N_{k+1}}
\cdots\left|W\right\rangle_{N_{n-2}}\left|0\rangle\right\rangle_{N_{n-1}}\nonumber\\
&&+\cdots\nonumber\\
&&+a_{2^{n-k}-1}\left|W\right\rangle_{N_k}\left|W\right\rangle_{N_{k+1}}
\cdots\left|W\right\rangle_{N_{n-2}}\left|W\right\rangle_{N_{n-1}})\nonumber\\
&&+
\left|0\rangle\right\rangle_{N_0}\left|0\rangle\right\rangle_{N_1}
\cdots\left|0\rangle\right\rangle_{N_{k-2}}\left|1\rangle\right\rangle_{N_{k-1}}\nonumber\\
&&(a_{2^{n-k}}\left|0\rangle\right\rangle_{N_k}\left|0\rangle\right\rangle_{N_{k+1}}
\cdots\left|0\rangle\right\rangle_{N_{n-2}}\left|0\rangle\right\rangle_{N_{n-1}}\nonumber\\
&&+a_{2^{n-k}+1}\left|0\rangle\right\rangle_{N_k}\left|0\rangle\right\rangle_{N_{k+1}}
\cdots\left|0\rangle\right\rangle_{N_{n-2}}\left|W\right\rangle_{N_{n-1}}\nonumber\\
&&+a_{2^{n-k}+2}\left|0\rangle\right\rangle_{N_k}\left|0\rangle\right\rangle_{N_{k+1}}
\cdots\left|W\right\rangle_{N_{n-2}}\left|0\rangle\right\rangle_{N_{n-1}}\nonumber\\
&&+\cdots\nonumber\\
&&+a_{2\cdot2^{n-k}-1}\left|W\right\rangle_{N_k}\left|W\right\rangle_{N_{k+1}}
\cdots\left|W\right\rangle_{N_{n-2}}\left|W\right\rangle_{N_{n-1}})\nonumber
\end{eqnarray}
\begin{eqnarray}
&&+\cdots\nonumber\\
&&+
\left|1\rangle\right\rangle_{N_0}\left|1\rangle\right\rangle_{N_1}
\cdots\left|1\rangle\right\rangle_{N_{k-2}}\left|1\rangle\right\rangle_{N_{k-1}}\nonumber\\
&&(a_{(2^k-1)2^{n-k}}\left|0\rangle\right\rangle_{N_k}\left|0\rangle\right\rangle_{N_{k+1}}
\cdots\left|0\rangle\right\rangle_{N_{n-2}}\left|0\rangle\right\rangle_{N_{n-1}}\nonumber\\
&&+a_{(2^k-1)2^{n-k}+1}\left|0\rangle\right\rangle_{N_k}\left|0\rangle\right\rangle_{N_{k+1}}
\cdots\left|0\rangle\right\rangle_{N_{n-2}}\left|W\right\rangle_{N_{n-1}}\nonumber\\
&&+a_{(2^k-1)2^{n-k}+2}\left|0\rangle\right\rangle_{N_k}\left|0\rangle\right\rangle_{N_{k+1}}
\cdots\left|W\right\rangle_{N_{n-2}}\left|0\rangle\right\rangle_{N_{n-1}}\nonumber\\
&&+\cdots\nonumber\\
&&+a_{2^n-1}\left|W\right\rangle_{N_k}\left|W\right\rangle_{N_{k+1}}
\cdots\left|W\right\rangle_{N_{n-2}}\left|W\right\rangle_{N_{n-1}}),\nonumber\\
&&a_i\in \mathcal{C},i=0,1,...,2^n-1; N_i\geq2,i=0,1,...,n-1.\nonumber\\
\end{eqnarray}
That is, $\left|\Omega_{GHZ-W}\right\rangle_{k,n-k}$ represents
the most universal form composed of the GHZ-type basis and the
W-type basis, so we call it the GHZ-W-type state. In particular,
when $k=n$ the state is completely composed of the GHZ-type basis,
and define
$\left|\Omega_{GHZ}\right\rangle_n\equiv\left|\Omega_{GHZ-W}\right\rangle_{n,0}$,
which we call the GHZ-type state. Similarly, the W-type state is
defined as
$\left|\Omega_W\right\rangle_n\equiv\left|\Omega_{GHZ-W}\right\rangle_{0,n}$.
Although the GHZ-type and W-type states are merely two special
cases of $\left|\Omega_{GHZ-W}\right\rangle_{k,n-k}$, we can
character many entanglement properties of
$\left|\Omega_{GHZ-W}\right\rangle_{k,n-k}$ by investigating these
two representatives. In what follows we analyze the property of
these three types of states, including detecting the full
entanglement and the condition for the simplest form of the
GHZ-W-type state. Our interest mainly focuses on the
classification of this multiqubit entanglement. In particular, we
will classify the GHZ-type and W-type state in a regular way,
which proves to be related to the theory of enumerative
combinatorics \cite{Stanley}. We thus have converted the issue of
classification of this family of multiqubit state into a
combinatorial problem.\\
\\

\begin{center}
{\bf A. Detecting the entanglement of the GHZ-W-type state}
\end{center}

Let us first find out the condition on which the state
$\left|\Omega_{GHZ-W}\right\rangle_{k,n-k}$ is fully entangled,
since the separable state can always be expressed as a direct
product of several truly entangled states. It is easy to see that
the structure of $\left|\Omega_{GHZ-W}\right\rangle_{k,n-k}$
resembles that of the general pure multiqubit state,
\begin{eqnarray}
\left|\Omega\right\rangle_n&\equiv&b_0|\overbrace{0,0,...,0,0}^n\rangle
+b_1\left|0,0,...,0,1\right\rangle+\nonumber\\
&&b_2\left|0,0,...,1,0\right\rangle+\cdots+b_{2^n-1}\left|1,1,...,1,1\right\rangle.\nonumber\\
\end{eqnarray}
However, giving an exhaustive condition on which one can judge
whether $\left|\Omega\right\rangle_n$ is fully entangled is
difficult. Theoretically, one can do it by checking the rank of
each k-partite reduced matrix operator, $1\leq k\leq[n/2]$; that
is, $\left|\Omega\right\rangle_n$ is fully entangled iff each rank
is larger than 1. In this case, one has to check $2^{n-1}-1$
matrices in all. It actually concerns the separability problem
\cite{Doherty} and remains a puzzling aspect of QIT. Furthermore,
usually the state $\left|\Omega_{GHZ-W}\right\rangle_{k,n-k}$ is
not the locally symmetric extension (LSE) of some
$\left|\Omega\right\rangle_n$ \cite{Doherty}, for both of them are
pure. Nevertheless, this "pseudo" parallelism still helps detect
the entanglement of the GHZ-W-type state, an interesting relation
can be constructed.

\textit{Theorem 2}. Construct the corresponding state
$\left|\Omega\right\rangle_n$ of
$\left|\Omega_{GHZ-W}\right\rangle_{k,n-k}$, by ``concentrating"
the $i$'th group consisting of $N_i$ parties in
$\left|\Omega_{GHZ-W}\right\rangle_{k,n-k}$ to the $i$'th party in
$\left|\Omega\right\rangle_n,$ i.e., $
\left|0\rangle\right\rangle_{N_i}\rightarrow\left|0\right\rangle_i,$
$\left|1\rangle\right\rangle_{N_i}$ or
$\left|W\right\rangle_{N_i}\rightarrow\left|1\right\rangle_i.$
Then $\left|\Omega_{GHZ-W}\right\rangle_{k,n-k}$ is fully
entangled iff $\left|\Omega\right\rangle_n$ is fully entangled.

\textit{Proof}. The necessity could be lightly obtained by the
parallelism between $\left|\Omega_{GHZ-W}\right\rangle_{k,n-k}$
and $\left|\Omega\right\rangle_n$. For the case of sufficiency, we
can prove the assertion by reduction to absurdity. Regard
$\left|\Omega_{GHZ-W}\right\rangle_{k,n-k}$ as a bipartite state
by random partition, and the local rank of this bipartite state
must always be larger than 1 iff it is fully entangled. Suppose
the bipartition happens in the $i$'th group whose space is spanned
by $R_{N_i1}$,
\begin{eqnarray}
\left|\Omega_{GHZ-W}\right\rangle_{k,n-k}&=&
(\left|0\rangle\right\rangle_m\left|\psi_0\right\rangle
+\left|1\rangle\right\rangle_m\left|\psi_1\right\rangle)\nonumber\\
&&\otimes
(\left|0\rangle\right\rangle_{N_i-m}\left|\phi_0\right\rangle
+\left|1\rangle\right\rangle_{N_i-m}\left|\phi_1\right\rangle)\nonumber\\
&=&\left|0\rangle\right\rangle_m\left|0\rangle\right\rangle_{N_i-m}
\left|\psi_0\right\rangle\left|\phi_0\right\rangle+\nonumber\\
&&\left|0\rangle\right\rangle_m\left|1\rangle\right\rangle_{N_i-m}
\left|\psi_0\right\rangle\left|\phi_1\right\rangle+\nonumber\\
&&\left|1\rangle\right\rangle_m\left|0\rangle\right\rangle_{N_i-m}
\left|\psi_1\right\rangle\left|\phi_0\right\rangle+\nonumber\\
&&\left|1\rangle\right\rangle_m\left|1\rangle\right\rangle_{N_i-m}
\left|\psi_1\right\rangle\left|\phi_1\right\rangle,\nonumber\\
&&0<m<N_i.
\end{eqnarray}
If $\left|\psi_0\right\rangle$ and $\left|\psi_1\right\rangle$ are
linearly dependent, the space of $i$'th group is spanned by the
basis
$\{(\alpha\left|0\rangle\right\rangle_m+\beta\left|1\rangle\right\rangle_m)\left|0\rangle\right\rangle_{N_i-m},
(\alpha\left|0\rangle\right\rangle_m+\beta\left|1\rangle\right\rangle_m)\left|1\rangle\right\rangle_{N_i-m}\}$,
which evidently cannot be $R_{N_i0}$ or $R_{N_i1}$, so they must
be linearly independent. Similarly, $\left|\phi_0\right\rangle$
and $\left|\phi_1\right\rangle$ are also linearly independent and
thus the local rank of $i$'th group is 4, which is inconsistent
with the assumption that the corresponding state
$\left|\Omega\right\rangle_n$ is a truly entangled multiqubit
state. For the case in which bipartition happens in one group of
$R_{N0}$, the above argument entirely applies. Conclusively,
$\left|\Omega_{GHZ-W}\right\rangle_{k,n-k}$ is fully entangled iff
its corresponding state $\left|\Omega\right\rangle_n$ is fully
entangled. \hspace*{\fill}$\blacksquare$

Thus we can judge whether
$\left|\Omega_{GHZ-W}\right\rangle_{k,n-k}$ is fully entangled by
only analyzing its corresponding state
$\left|\Omega\right\rangle_n$. For example, the following state is
always fully entangled iff $\alpha\beta\gamma\neq0$,
\begin{eqnarray}
\left|\Psi\right\rangle&=&\alpha\left|000\right\rangle\left|000\right\rangle
(\left|001\right\rangle+\left|001\right\rangle+\left|100\right\rangle)\nonumber\\
&&+\beta\left|000\right\rangle\left|111\right\rangle\left|000\right\rangle
+\gamma\left|111\right\rangle\left|000\right\rangle\left|000\right\rangle,
\end{eqnarray}
whose corresponding state is
$\alpha\left|001\right\rangle+\beta\left|010\right\rangle+\gamma\left|100\right\rangle$.
In this sense, theorem 2 efficiently reduces the calculation
required in entanglement detection. As mentioned above, one has to
calculate $2^{\sum^{n-1}_{i=0}N_i-1}-1$ matrices to judge whether
a pure $(\sum^{n-1}_{i=0}N_i)$-qubit state is fully entangled.
However, if it can be structurally adjusted to fit the form of
GHZ-W-type state, then it suffices to calculate $2^{n-1}-1$
matrices in all. To make for a better effect in the following
text, we propose an important property of pure multiqubit states.
Though detecting the multiqubit entanglement is generally
difficult, it is shown to be much easier to determine whether a
multiqubit state with undetermined coefficients $could$ be fully
entangled by choosing proper coefficients. This helps characterize
the multiqubit entanglement.

\textit{Theorem 3}. Suppose
$\{\left|i_0,i_1,...,i_{n-1}\right\rangle,i_k=0,1,k=0,1,...,n-1\}$
is the set of basis vectors on the space ${\cal H}_{P_0,P_1,\cdots
P_{n-1}}$ and choose $k\geq2$ vectors
$\left|\phi_j\right\rangle=|i^j_0,i^j_1,...,i^j_{n-1}\rangle,j=0,1,...,k-1$
from this set. Then there exist infinitely many sets of
coefficients $\{a_0,a_1,...,a_{k-1}\}$, such that
$\sum^{k-1}_{i=0} a_i\left|\phi_i\right\rangle$ is fully entangled
iff there exist at least two inequivalent states in any set
$\{\left|i^j_s\right\rangle,j=0,1,...,k-1,s\in[0,n-1]\}$. That is,
$\left|i^j_s\right\rangle$'s $,j=0,1,...,k-1$ cannot
simultaneously be $\left|0\right\rangle$ or
$\left|1\right\rangle$.

\textit{Proof}. We only need to verify the sufficiency. Let
$\left|\Psi\right\rangle_{P_0,P_1,\cdots P_{n-1}}=\sum^{k-2}_{i=0}
a_i\left|\phi_i\right\rangle+x\left|\phi_{k-1}\right\rangle,x=a_{k-1}$.
Assume that $x,a_i>0,i=0,1,...,k-2$, since it suffices to prove
the statement in this case. For the first step, consider the
1-party reduced matrix operator of $P_0$ (possibly after
performing a permutation matrix),
\begin{eqnarray}
\rho^{P_0}&=&(b\left|0\right\rangle+x\left|1\right\rangle)
(b\left\langle0\right|+x\left\langle1\right|)\nonumber\\
&&+\sum_i(c_i\left|0\right\rangle+d_i\left|1\right\rangle)
(c_i\left\langle0\right|+d_i\left\langle1\right|),
\end{eqnarray}
and let $A=\sum_ic^2_i,B=\sum_ic_id_i,C=\sum_id^2_i$, then\\
$\rho^{P_0}=\left(\begin{array}{cc}
A+b^2 & B+bx \\
B+bx & C+x^2
\end{array}\right)$, det$\rho^{P_0}=Ax^2-2Bbx+AC+b^2C-B^2$.
Clearly, $P_0$ is separate from the other parties iff
det$\rho^{P_0}=0$. Let us consider the condition making
det$\rho^{P_0}\equiv0$, i.e., no matter what $x$ equals. First, it
requires $A=0$, which means $c_i=0$ for each $i$. Thus, $B=0$ and
$b>0$ for there exist at least one vector
$\left|0\right\rangle_{P_0}$, which also implies $C>0$. In this
case, det$\rho^{P_0}=b^2C>0$. Consequently, it is not possible to
realize det$\rho^{P_0}\equiv0$, so there is only a finite number
of $x$'s (at most two here) making det$\rho^{P_0}=0$. Construct a
set $S_0$, such that $x\in S_0$ making det$\rho^{P_0}=0$.
Similarly, we can construct the sets $S_1,S_2,...,S_{n-1}$ for the
parties $P_1,P_2,...,P_{n-1}$ respectively. Subsequently for the
second step, let us move to the case of 2-party reduced matrix
operator of $P_0P_1$,
\begin{eqnarray}
\rho^{P_0P_1}&=&(a\left|00\right\rangle+b\left|01\right\rangle+c\left|10\right\rangle+x\left|11\right\rangle)\nonumber\\
&&(a\left\langle00\right|+b\left\langle01\right|+c\left\langle10\right|+x\left\langle11\right|)\nonumber\\
&&+\sum_i(d_i\left|00\right\rangle+e_i\left|01\right\rangle+f_i\left|10\right\rangle+g_i\left|11\right\rangle)\nonumber\\
&&(d_i\left\langle00\right|+e_i\left\langle01\right|+f_i\left\langle10\right|+g_i\left\langle11\right|),
\end{eqnarray}
which could be rewritten as \\
$\rho^{P_0P_1}=\left(\begin{array}{cccc}
a^2 & ab & ac & ax\\
ba & b^2 & bc & bx\\
ca & cb & c^2 & cx\\
xa & xb & xc & x^2\\
\end{array}\right)+
\left(\begin{array}{cccc}
A_{00} & A_{01} & A_{02} & A_{03}\\
A_{10} & A_{11} & A_{12} & A_{13}\\
A_{20} & A_{21} & A_{22} & A_{23}\\
A_{30} & A_{31} & A_{32} & A_{33}\\
\end{array}\right)$.\\
Consider the principal submatrices of $\rho^{P_0P_1}$,\\
$M_0=\left(\begin{array}{cc}
A_{00}+a^2 & A_{03}+ax \\
A_{30}+ax & A_{33}+x^2
\end{array}\right)$,\\
$M_1=\left(\begin{array}{cc}
A_{11}+b^2 & A_{13}+bx \\
A_{31}+bx & A_{33}+x^2
\end{array}\right)$,\\
and\\
$M_2=\left(\begin{array}{cc}
A_{22}+c^2 & A_{23}+cx \\
A_{32}+cx & A_{33}+x^2
\end{array}\right)$.\\
Notice that $P_0,P_1$ are always separate from the other parties
iff rank$[\rho^{P_0P_1}]\equiv1$, which requires
det${M_i}\equiv0,i=0,1,2$. Applying the result in the first step
to these matrices leads to $A_{00}=A_{11}=A_{22}=0$, which implies
there is at least one nonzero number in $a,b,c$, as well as
$A_{33}>0$ based on the reason similar to that in the first step.
Without loss of generality, choose $a>0$ and
det${M_0}=a^2A_{33}>0$. So it is also impossible to make
rank$[\rho^{P_0P_1}]\equiv1$. Again we construct a finite set
$S_{01}$, such that $x\in S_{01}$ making $[\rho^{P_0P_1}]\equiv1$.
Similarly, we can construct the sets
$S_{02},S_{12},S_{13},...,S_{n-2,n-1}$ for the subgroups
$P_0P_2,P_1P_2,P_1P_3,...,P_{n-2}P_{n-1}$ respectively. Continue
until all reduced matrix operators have been checked and exhibit
the set
\begin{eqnarray}
S&=&S_0\sqcup S_1\sqcup\cdots S_{n-1}\sqcup S_{01}\sqcup
S_{02}\sqcup S_{12}\sqcup\cdots\sqcup \nonumber\\
&&S_{n-2,n-1}\sqcup\cdots\sqcup S_{0,1,2,...n-3,n-2}\sqcup S_{0,1,2,...n-3,n-1}\nonumber\\
&&\sqcup S_{0,1,2,...,n-4,n-2,n-1}\sqcup\cdots\sqcup
S_{1,2,3,...,n-3,n-2,n-1},\nonumber\\
\end{eqnarray}
which contains a finite amount of (complex) numbers. Choose
$x\not\in S>0$, then the set of numbers $\{a_0,a_1,...,a_{k-1}\}$
makes $\sum^{k-1}_{i=0} a_i\left|\phi_i\right\rangle$ fully
entangled. Since we can randomly choose $a_0,a_1,...,a_{k-2}>0$,
there exist infinitely many sets of numbers satisfying the
statement. This completes the proof. \hspace*{\fill}$\blacksquare$

This result gives a useful criterion for determining whether a
multiqubit state can be fully entangled, especially in the cases
containing a great number of qubits. By virtue of theorem 3, we
say that a general multiqubit state can always be fully entangled
and we have to classify it, if no range of its single reduced
matrix operator merely contains the same computational basis
$\left|0\right\rangle$ or $\left|1\right\rangle$. Without loss of
generality, we suppose that the states in the following text
always satisfies this condition and are supposed to be fully
entangled, e.g., $\left|\Psi\right\rangle=
\alpha\left|00\right\rangle(\left|01\right\rangle+\left|10\right\rangle)
+\beta\left|11\right\rangle\left|00\right\rangle $.\\
\\

\begin{center}
{\bf B. Simplest form of GHZ-W-type state}
\end{center}

In this subsection, we investigate the condition under which the
GHZ-W-type state will be of the simplest form defined below. Like
the condition of full entanglement in section A, the simplest form
is also necessary when classifying the GHZ-W-type multiqubit
entanglement.

Let
\begin{eqnarray}
\left|\Omega_{GHZ-W}\right\rangle_{k,n-k}=\sum_i
c_i\left|a_{i,0}\rangle\right\rangle_{N_0}\left|a_{i,1}\rangle\right\rangle_{N_1}
\cdots\left|a_{i,k-1}\rangle\right\rangle_{N_{k-1}}\nonumber\\
\left|a_{i,k}\right\rangle_{N_k}\left|a_{i,k+1}\right\rangle_{N_{k+1}}
\cdots\left|a_{i,n-1}\right\rangle_{N_{n-1}},\nonumber\\
\end{eqnarray}
where $a_{i,j}=0,1$ for $j=0,...,k-1$ and $a_{i,j}=0,W$ for
$j=k,...,n-1$. Due to the definition of GHZ-W-type state, the
space of the i'th group consisting of $N_i$ particles is spanned
by $R_{N_i0}$ or $R_{N_i1}$. Interestingly, there may exist
several groups containing $N_i$ particles respectively, which will
be absorbed into a larger group containing $\sum_iN_i$ particles
and the space of this new group is spanned by $R_{\sum_iN_i0}$ or
$R_{\sum_iN_i1}$. So we can collect these groups respectively to
get a more brief expression of the state
$\left|\Omega_{GHZ-W}\right\rangle_{k,n-k}$. Continue this
procedure and finally we can get
\begin{eqnarray}
&\left|\Omega_{GHZ-W}\right\rangle_{k^{\prime},n^{\prime}-k^{\prime}}=&\sum_{i^{\prime}}
c_{i^{\prime}}\left|a_{i^{\prime},0}\rangle\right\rangle_{N_0}\left|a_{i^{\prime},1}\rangle\right\rangle_{N_1}
\cdots\nonumber
\\&\left|a_{i,k^{\prime}-1}\rangle\right\rangle_{N_{k^{\prime}-1}}
\left|a_{i^{\prime},k^{\prime}}\right\rangle_{N_{k^{\prime}}}&
\left|a_{i^{\prime},k^{\prime}+1}\right\rangle_{N_{k^{\prime}+1}}
\cdots\left|a_{i,n^{\prime}-1}\right\rangle_{N_{n^{\prime}-1}},\nonumber\\
\end{eqnarray}
which is a succinct expression of
$\left|\Omega_{GHZ-W}\right\rangle_{k^{\prime},n^{\prime}-k^{\prime}}$
containing $n^{\prime}\leq n$ groups. In this state, the space of
the i'th group consisting of $N^{\prime}_i$ particles is spanned
by $R_{N^{\prime}_i0}$ or $R_{N^{\prime}_i1}$, and no two groups
will be absorbed into a larger group whose space is still spanned
by the GHZ-type or W-type basis. We then call the above expression
the $simplest$ form of $\left|\Omega_{GHZ-W}\right\rangle_{k,n-k}$
(or this expression is simplest). For instance, the following
state is in the simplest form,
\begin{equation}
\left|\Psi\right\rangle=
\left|0\rangle\right\rangle_{N_0}\left|0\rangle\right\rangle_{N_1}+
\left|1\rangle\right\rangle_{N_0}\left|W\right\rangle_{N_1},
\end{equation}
where the two groups here cannot form a larger group whose space
is spanned by the GHZ-type or W-type basis. However, it is not
easy to check whether a general GHZ-W-type state is under the
simplest form, especially for those consisting of many groups.
Consider a 3-group W-type state
\begin{eqnarray}
\left|\Psi^{\prime}\right\rangle&=&
\left|0\rangle\right\rangle_{N_0}\left|0\rangle\right\rangle_{N_1}\left|0\rangle\right\rangle_{N_2}+
\alpha\left|W\right\rangle_{N_0}\left|W\right\rangle_{N_1}\left|0\rangle\right\rangle_{N_2}\nonumber\\
&+&\beta\left|0\rangle\right\rangle_{N_0}\left|W\right\rangle_{N_1}\left|W\right\rangle_{N_2},
\alpha\beta\neq0.
\end{eqnarray}
By theorem 2 and 3, it is fully entangled. Notice the coefficients
$\alpha$ and $\beta$ can be moved away by the ILO's
$\otimes\prod^{N_0-1}_{i=0}O^i_1(\left|1\right\rangle,1/\alpha)$
and
$\otimes\prod^{N_2-1}_{i=0}O^{i+N_0+N_1}_1(\left|1\right\rangle,1/\beta)$
on the first $N_0$ and last $N_2$ parties respectively. Now, one
can find out that the 0'th and 2'th groups can form a larger group
whose space is spanned by the W-type basis,
\begin{eqnarray}
\left|\Psi^{\prime}\right\rangle&\sim&
\left|0\rangle\right\rangle_{N_1}\left|0\rangle\right\rangle_{N_0}\left|0\rangle\right\rangle_{N_2}\nonumber\\
&+&\left|W\right\rangle_{N_1}(\left|W\right\rangle_{N_0}\left|0\rangle\right\rangle_{N_2}
+\left|0\right\rangle_{N_0}\left|W\right\rangle_{N_2})\nonumber\\
&=&\left|0\rangle\right\rangle_{N_1}\left|0\rangle\right\rangle_{N_0+N_2}
+\left|W\right\rangle_{N_1}\left|W\right\rangle_{N_0+N_2}.
\end{eqnarray}
Hence, this expression is actually the simplest form of
$\left|\Psi^{\prime}\right\rangle$. Similar mergers exist in the
case of GHZ-type state. For example, we prove theorem 2 by
disproval where the bipartition makes a group of partners whose
space is spanned by the GHZ-type basis into two groups, while
either of them is in the space spanned by the GHZ-type basis.

To characterize the entangled family
$\left|\Omega_{GHZ-W}\right\rangle_{k,n-k}$, we must find out the
simplest form of any state of it. The reason derives from that the
local rank is invariant under the ILO's. By the reset of the
positions of the particles in the system, one can merge several
groups into a new group consisting of more particles whose space
is spanned by the same SEB, i.e., the local rank of the new group
is identical to that of either of the subgroups in it. In this
case, it is hard to decide the correspondence of the initial
particles and the final ones in the system, just similar to that
any local rank is 2 in the multiqubit state. Consequently, a
theoretical depiction is required for finding out the simplest
form of $\left|\Omega_{GHZ-W}\right\rangle_{k,n-k}$. Rewrite it as
follows,
\begin{eqnarray}
\left|\Omega_{GHZ-W}\right\rangle_{k,n-k}&=&
\left|0\rangle\right\rangle_{N_0}\left|0\rangle\right\rangle_{N_1}\left|\psi_0\right\rangle\nonumber\\
&&+\left|0\rangle\right\rangle_{N_0}\left|1\rangle\right\rangle_{N_1}\left|\psi_1\right\rangle\nonumber\\
&&+\left|1\rangle\right\rangle_{N_0}\left|0\rangle\right\rangle_{N_1}\left|\psi_2\right\rangle\nonumber\\
&&+\left|1\rangle\right\rangle_{N_0}\left|1\rangle\right\rangle_{N_1}\left|\psi_3\right\rangle\nonumber\\
&=&\left|00\right\rangle_{AB}(\left|0\rangle\right\rangle_{N_0-1}
\left|0\rangle\right\rangle_{N_1-1}\left|\psi_0\right\rangle)_C\nonumber\\
&&\left|01\right\rangle_{AB}(\left|0\rangle\right\rangle_{N_0-1}
\left|1\rangle\right\rangle_{N_1-1}\left|\psi_1\right\rangle)_C\nonumber\\
&&\left|10\right\rangle_{AB}(\left|1\rangle\right\rangle_{N_0-1}
\left|0\rangle\right\rangle_{N_1-1}\left|\psi_2\right\rangle)_C\nonumber\\
&&\left|11\right\rangle_{AB}(\left|1\rangle\right\rangle_{N_0-1}
\left|1\rangle\right\rangle_{N_1-1}\left|\psi_3\right\rangle)_C,\nonumber\\
\end{eqnarray}
notice $N_0,N_1\geq2$ here. In the light of theorem 1, a space is
spanned by $R_{N0}$ or $R_{N1}$ iff any two-qubit subspace of it
is spanned by $R_{20}$ or $R_{21}$. This implies that if the
$0$'th and $1$'th group can be merged into one group, the $AB$
space must be spanned by $R_{21}$. However, the four nonvanishing
vectors of system $C$ form a set of orthogonal bases. Thus, the
two groups make up a new group iff
$\left|\psi_0\right\rangle=\left|\psi_3\right\rangle=0$, or
$\left|\psi_1\right\rangle=\left|\psi_2\right\rangle=0$.

On the other hand, the GHZ-W-type state can also be expressed in
terms of the W-type basis,
\begin{eqnarray}
\left|\Omega_{GHZ-W}\right\rangle_{k,n-k}&=&
\left|0\rangle\right\rangle_{N_0}\left|0\rangle\right\rangle_{N_1}\left|\psi_0\right\rangle\nonumber\\
&&+\left|0\rangle\right\rangle_{N_0}\left|W\right\rangle_{N_1}\left|\psi_1\right\rangle\nonumber\\
&&+\left|W\right\rangle_{N_0}\left|0\rangle\right\rangle_{N_1}\left|\psi_2\right\rangle\nonumber\\
&&+\left|W\right\rangle_{N_0}\left|W\right\rangle_{N_1}\left|\psi_3\right\rangle\nonumber\\
&=&\left|00\right\rangle_{AB}
(\left|0\rangle\right\rangle_{N_0-1}\left|0\rangle\right\rangle_{N_1-1}\left|\psi_0\right\rangle\nonumber\\
&&+\left|0\rangle\right\rangle_{N_0-1}\left|W\right\rangle_{N_1-1}\left|\psi_1\right\rangle\nonumber\\
&&+\left|W\right\rangle_{N_0-1}\left|0\rangle\right\rangle_{N_1-1}\left|\psi_2\right\rangle\nonumber\\
&&+\left|W\right\rangle_{N_0-1}\left|W\right\rangle_{N_1-1}\left|\psi_3\right\rangle)_C\nonumber\\
&&+\left|01\right\rangle_{AB}
(\left|0\rangle\right\rangle_{N_0-1}\left|0\rangle\right\rangle_{N_1-1}\left|\psi_1\right\rangle\nonumber\\
&&+\left|W\right\rangle_{N_0-1}\left|0\rangle\right\rangle_{N_1-1}\left|\psi_3\right\rangle)_C\nonumber\\
&&+\left|10\right\rangle_{AB}
(\left|0\rangle\right\rangle_{N_0-1}\left|0\rangle\right\rangle_{N_1-1}\left|\psi_2\right\rangle\nonumber\\
&&+\left|0\rangle\right\rangle_{N_0-1}\left|W\right\rangle_{N_1-1}\left|\psi_3\right\rangle)_C\nonumber\\
&&+\left|11\right\rangle_{AB}
(\left|0\rangle\right\rangle_{N_0-1}\left|0\rangle\right\rangle_{N_1-1}\left|\psi_3\right\rangle)_C.\nonumber\\
\end{eqnarray}
Due to theorem 1, the $AB$ space should be spanned by $R_{20}$. If
$\left|\psi_3\right\rangle\neq0$, one can easily check that the
four vectors of system $C$ are linearly independent. Since the
rank of the $AB$ system should be equal to that of the $C$ system,
there must be $\left|\psi_3\right\rangle=0$, i.e.,
\begin{eqnarray}
&\alpha&(\left|0\rangle\right\rangle_{N_0-1}\left|0\rangle\right\rangle_{N_1-1}\left|\psi_0\right\rangle+
\left|0\rangle\right\rangle_{N_0-1}\left|W\right\rangle_{N_1-1}\left|\psi_1\right\rangle\nonumber\\
&+&\left|W\right\rangle_{N_0-1}\left|0\rangle\right\rangle_{N_1-1}\left|\psi_2\right\rangle)\nonumber\\
&+&\beta\left|0\rangle\right\rangle_{N_0-1}\left|0\rangle\right\rangle_{N_1-1}\left|\psi_1\right\rangle\nonumber\\
&+&\gamma\left|0\rangle\right\rangle_{N_0-1}\left|0\rangle\right\rangle_{N_1-1}\left|\psi_2\right\rangle\nonumber\\
&=&0,
\end{eqnarray}
which leads to
$\alpha=0,\beta\left|\psi_1\right\rangle=-\gamma\left|\psi_2\right\rangle\neq0$.
Combined with the result of GHZ-type state, we have

\textit{Corollary 1.} Given a fully entangled state
$\left|\Omega_{GHZ-W}\right\rangle_{k,n-k}$, it is in the simplest
form iff $\left|\psi_0\right\rangle=\left|\psi_3\right\rangle=0$,
or $\left|\psi_1\right\rangle=\left|\psi_2\right\rangle=0$ when
$\left|\Omega_{GHZ-W}\right\rangle_{k,n-k}=
\left|0\rangle\right\rangle_{N_0}\left|0\rangle\right\rangle_{N_1}\left|\psi_0\right\rangle
+\left|0\rangle\right\rangle_{N_0}\left|1\rangle\right\rangle_{N_1}\left|\psi_1\right\rangle
+\left|1\rangle\right\rangle_{N_0}\left|0\rangle\right\rangle_{N_1}\left|\psi_2\right\rangle
+\left|1\rangle\right\rangle_{N_0}\left|1\rangle\right\rangle_{N_1}\left|\psi_3\right\rangle
$; or iff
$\left|\psi_1\right\rangle=k\left|\psi_2\right\rangle,k\neq0$ and
$\left|\psi_3\right\rangle=0$ when
$\left|\Omega_{GHZ-W}\right\rangle_{k,n-k}=
\left|0\rangle\right\rangle_{N_0}\left|0\rangle\right\rangle_{N_1}\left|\psi_0\right\rangle
+\left|0\rangle\right\rangle_{N_0}\left|W\right\rangle_{N_1}\left|\psi_1\right\rangle
+\left|W\right\rangle_{N_0}\left|0\rangle\right\rangle_{N_1}\left|\psi_2\right\rangle+\\
\left|W\right\rangle_{N_0}\left|W\right\rangle_{N_1}\left|\psi_3\right\rangle$.
We call these two conditions GHZ-criterion and W-criterion
respectively. Besides, the merge will not happen between a
GHZ-type basis and W-type basis. \hspace*{\fill}$\blacksquare$

So we have obtained the theoretical method to find out the
simplest form of a general GHZ-W-type state. Practically, one has
to consider each group whose space contains the GHZ-type and
W-type basis respectively. Even so, it is not difficult to
determine whether a state is under the simplest form or not. For
example, consider the following state,
\begin{eqnarray}
\left|\Psi\right\rangle&=&
a_0\left|0\rangle\right\rangle_{N_0}\left|0\rangle\right\rangle_{N_1}\left|0\rangle\right\rangle_{N_2}
\left|0\rangle\right\rangle_{N_3}\left|0\rangle\right\rangle_{N_4}\left|0\rangle\right\rangle_{N_5}+\nonumber\\
&&a_1\left|0\rangle\right\rangle_{N_0}\left|0\rangle\right\rangle_{N_1}\left|1\rangle\right\rangle_{N_2}
\left|W\right\rangle_{N_3}\left|0\rangle\right\rangle_{N_4}\left|0\rangle\right\rangle_{N_5}+\nonumber\\
&&a_2\left|0\rangle\right\rangle_{N_0}\left|1\rangle\right\rangle_{N_1}\left|0\rangle\right\rangle_{N_2}
\left|0\rangle\right\rangle_{N_3}\left|W\right\rangle_{N_4}\left|0\rangle\right\rangle_{N_5}+\nonumber\\
&&a_3\left|1\rangle\right\rangle_{N_0}\left|0\rangle\right\rangle_{N_1}\left|0\rangle\right\rangle_{N_2}
\left|0\rangle\right\rangle_{N_3}\left|0\rangle\right\rangle_{N_4}\left|W\right\rangle_{N_5},\nonumber\\
\end{eqnarray}
which is fully entangled by theorem 3. Notice the state is partly
symmetric, i.e., unchanged under the exchange of groups 0 and 1, 4
and 5 simultaneously, etc. We analyze the part of GHZ-type basis
and W-type basis respectively. Since there always exist three kind
of bases in each group's space, i.e.,
$\left|0\rangle\right\rangle_{N_i}\left|0\rangle\right\rangle_{N_j},
\left|0\rangle\right\rangle_{N_i}\left|1\rangle\right\rangle_{N_j},
\left|1\rangle\right\rangle_{N_i}\left|0\rangle\right\rangle_{N_j},$
so the part of GHZ-type basis is under the simplest form. While
for the case of another part, observe that, e.g., the states
$\left|0\rangle\right\rangle_{N_0}\left|1\rangle\right\rangle_{N_1}\left|0\rangle\right\rangle_{N_2}
\left|0\rangle\right\rangle_{N_3}$ and
$\left|1\rangle\right\rangle_{N_0}\left|0\rangle\right\rangle_{N_1}\left|0\rangle\right\rangle_{N_2}
\left|0\rangle\right\rangle_{N_3}$ are orthogonal, which means the
part of W-type basis is also under the simplest form.
Conclusively, the state $\left|\Psi\right\rangle$ is under the
simplest form.

In general, an arbitrary GHZ-W-type state can always be turned
into its simplest form, which is an effective constitution at our
disposal. We then catalog
$\left|\Omega_{GHZ-W}\right\rangle_{k,n-k}$ under its simplest
form. First, two states
$\left|\Omega^A_{GHZ-W}\right\rangle_{k_0,n_0-k_0}$ and
$\left|\Omega^B_{GHZ-W}\right\rangle_{k_1,n_1-k_1}$ can be
classified into the same sort iff $n_0=n_1=n,k_0=k_1$,
$N^A_i=N^B_i,i=0,1,...,n-1$, up to some permutation of the
subgroups in the state. This is a necessary step in the process of
classification, and we will concretely describe how to write out
the inequivalent classes of entanglement under SLOCC with definite
$k_0,n_0$ in the next section. Second, from the above assertion,
it is natural to ask there exist how many ways of partition in
which an N-qubit GHZ-W-type state can be divided. To exemplify it,
a 6-qubit GHZ-type state can be in the following bipartite forms,
\begin{eqnarray}
\left|\Omega^A_{GHZ}\right\rangle_6&=&a_0\left|00\right\rangle\left|0000\right\rangle+
a_1\left|00\right\rangle\left|1111\right\rangle+\nonumber\\
&&a_2\left|11\right\rangle\left|0000\right\rangle+
a_3\left|11\right\rangle\left|1111\right\rangle,\nonumber\\
\left|\Omega^B_{GHZ}\right\rangle_6&=&b_0\left|000\right\rangle\left|000\right\rangle+
b_1\left|000\right\rangle\left|111\right\rangle+\nonumber\\
&&b_2\left|111\right\rangle\left|000\right\rangle+
b_3\left|111\right\rangle\left|111\right\rangle,
\end{eqnarray}
since $6=2+4=3+3$. However, one may notice there exists the third
way of partition for $6=2+2+2$, i.e.,
\begin{eqnarray}
\left|\Omega^C_{GHZ}\right\rangle_6&=&c_0\left|00\right\rangle\left|00\right\rangle\left|00\right\rangle+
c_1\left|00\right\rangle\left|00\right\rangle\left|11\right\rangle+\nonumber\\
&&c_2\left|00\right\rangle\left|11\right\rangle\left|00\right\rangle+
c_3\left|00\right\rangle\left|11\right\rangle\left|11\right\rangle+\nonumber\\
&&c_4\left|11\right\rangle\left|00\right\rangle\left|00\right\rangle+
c_5\left|11\right\rangle\left|00\right\rangle\left|11\right\rangle+\nonumber\\
&&c_6\left|11\right\rangle\left|11\right\rangle\left|00\right\rangle+
c_7\left|11\right\rangle\left|11\right\rangle\left|11\right\rangle.
\end{eqnarray}
The three states are in the simplest form by GHZ-criterion. So
there are three main classes of states in
$\left|\Omega_{GHZ}\right\rangle_6$, though either of which still
requires a more detailed analysis. For a natural number $N\geq4$,
first we divide it into two parts $M_{GHZ}$ and $M_W$,
representing the numbers of GHZ-type and W-type basis in the state
respectively. Mathematically, it is a problem of ordered partition
and the result is $N-3$ (notice $N_i\geq2$ for every group in
$\left|\Omega_{GHZ-W}\right\rangle_{k,n-k}$). Furthermore, one can
consider the ways of partition in $M_{GHZ}$ or $M_W$. This is a
problem of unordered partition and can be solved by the Ferrers
diagram \cite{Stanley}. Suppose the generating function is
\begin{equation}
G(y)_N\equiv(\sum^{\infty}_{i=0}y^i)(\sum^{\infty}_{i=0}y^{2i})\cdots(\sum^{\infty}_{i=0}y^{Ni}),
\end{equation}
and the coefficient of $y^m$ is $C[G(y)_N]_m$. Then the results
are $C[G(y)_{M_{GHZ}}]_{M_{GHZ}}-C[G(y)_{M_{GHZ}-1}]_{M_{GHZ}-1}$
for the partition of GHZ-type basis and
$C[G(y)_{M_W}]_{M_W}-C[G(y)_{M_W-1}]_{M_W-1}$ for that of W-type
basis, where $M_{GHZ}=N-M_W\in\{0\}\cup[2,N-2]\cup\{N\}$. Taking
one with another, a general GHZ-W-type state can be divided into
$L$ kinds of partitions, where
\begin{eqnarray}
L&=&\sum^{N-2}_{k=2}(C[G(y)_k]_k-C[G(y)_{k-1}]_{k-1})\nonumber\\
&\times&(C[G(y)_{N-k}]_{N-k}-C[G(y)_{N-k-1}]_{N-k-1})\nonumber\\
&+&2(C[G(y)_N]_N-C[G(y)_{N-1}]_{N-1}).
\end{eqnarray}
Consequently, by virtue of the theory of unordered partition one
can concretely write out all distinct ways in which $N$ can be
divided, and ulteriorly find out all main classes of states
$\left|\Omega_{GHZ-W}\right\rangle_{k,n-k}$. Since the
permutations between the subgroups have been considered by
unordered partition, so our deduction will not lead to the
repetition of classification.

\section{Classifying the GHZ-type state}

So far we managed to roughly character the structure of a general
GHZ-W-type state. In what follows,
$\left|\Omega_{GHZ-W}\right\rangle_{k,n-k}$ is always supposed to
be fully entangled and under the simplest form. As mentioned in
the preceding section, any N-qubit GHZ-W-type state can be divided
into many main classes of entanglement in terms of the theory of
unordered partition, and we now work on every class of state
particularly. Before moving to the general argument, it is a wise
choice to analyze the GHZ-type and W-type state respectively, for
they can be seen as two fundamental parts of a GHZ-W-type state.
In this section we character the GHZ-type state. In particular,
some important concepts are exhibited such as the relative ILO's
and columns, as well as the technique of enumerative
combinatorics. These tricks can be generalized to the case of the
W-type state, etc.

As for the SLOCC-classification, it is necessary to clarify the
form of the allowed ILO's operated by the partners in the system.
Consider the form of ILO
$V_{0,1,...,N-1}=\bigotimes\prod^{N-1}_{i=0}V_i$ making
$R_{N1}\rightarrow R_{N1}$ on the space ${\cal H}_{P_0,P_1,\cdots
P_{N-1}}$, where
\begin{eqnarray}
V_i=\left(\begin{array}{cc}
a_i & b_i \\
c_i & d_i
\end{array}\right)_{P_i},i=0,1,...,N-1,
\end{eqnarray}
and we have
\begin{eqnarray}
&V_{0,1,...,N-1}&(\alpha_0\left|0_0,0_1,...,0_{N-1}\right\rangle
+\alpha_1\left|1_0,1_1,...,1_{N-1}\right\rangle)\nonumber\\
&=&\beta_0\left|0_0,0_1,...,0_{N-1}\right\rangle+\beta_1\left|1_0,1_1,...,1_{N-1}\right\rangle,\nonumber\\
&\alpha_0,\alpha_1,\beta_0,\beta_1&\in\mathcal{C}.
\end{eqnarray}
Hence, there must be $b_i=c_i=0$ or $a_i=d_i=0,i=0,1,...,N-1$. In
other words, the $relative$
ILO can be expressed as \\
$V_{0,1,...,N-1}=\bigotimes\prod^{N-1}_{i=0}\left(\begin{array}{cc}
a_i & 0 \\
0 & 1
\end{array}\right)_{P_i},\forall a_i\neq0$, or \\
$V_{0,1,...,N-1}=\bigotimes\prod^{N-1}_{i=0}\left(\begin{array}{cc}
0 & b_i \\
1 & 0
\end{array}\right)_{P_i},\forall b_i\neq0$.\\
The total effect of this operator on the system is
\begin{equation}
V_{0,1,...,N-1}\left|0\rangle\right\rangle_N=\left|0\rangle\right\rangle_N,
V_{0,1,...,N-1}\left|1\rangle\right\rangle_N=\left|1\rangle\right\rangle_N,
\end{equation}
or
\begin{equation}
V_{0,1,...,N-1}\left|0\rangle\right\rangle_N=\left|1\rangle\right\rangle_N,
V_{0,1,...,N-1}\left|1\rangle\right\rangle_N=\left|0\rangle\right\rangle_N.
\end{equation}
In both cases, the universal factors have been omitted. Therefore,
the relative ILO only can realize the exchange of ``0" and ``1",
or they remain invariant. The essential change occurs in the
aspect of coefficients, by the optionally nonvanishing set
$\{a_0,a_1,...,a_{N-1}\}$ and $\{b_0,b_1,...,b_{N-1}\}$. Regard
each product
$\bigotimes\prod^{n-1}_{i=0}\left|x_i\rangle\right\rangle_{N_i},x_i=0$
or 1, as a term in the state $\left|\Omega_{GHZ}\right\rangle_n$,
then the ILO $V_{0,1,...,N-1}$ does not change the $number$ of
terms. That is,

\textit{Corollary 2}. Suppose two GHZ-type states are equivalent
under SLOCC, then they have the same number of terms.
\hspace*{\fill}$\blacksquare$

By virtue of this corollary, we can describe how to write out the
concrete form of the GHZ-type state. For convenience, let
$\bigotimes\prod^{n-1}_{i=0}\left|x_i\rangle\right\rangle_{N_i}=
\left|x_0,x_1,...,x_{n-1}\rangle\right\rangle$. That is, we
firstly find out all sorts of entanglement under the assumption
that the position of each group is ``equivalent", or under an
average division $N_i=N_j,i,j=0,1,...,n-1$. This means that we can
arbitrarily exchange the groups in the state. Subsequently, one
can permute the true sequence of every sort, $N_0,N_1,...,N_{n-1}$
for the groups in the state to generate a series of states. By
comparing these states and eliminating the repeating cases, all
residual states are indeed inequivalent. Finally, we should also
use the ILO's to adjust the coefficient of every term. Of all
above, we will mostly emphasize the first step for the posterior
two steps are more trivial. Some examples will be given to
illustrate them.

Write out the general form of the GHZ-type state,
\begin{eqnarray}
\left|\Omega_{GHZ}\right\rangle_n&=&a_0\left|0,...,0,0\rangle\right\rangle+
a_1\left|0,...,0,1\rangle\right\rangle+\cdots\nonumber\\
&+&a_{2^n-1}\left|1,...,1,1\rangle\right\rangle.
\end{eqnarray}
For the first step, let us start by an identification of the
following state,
\begin{eqnarray}
\left|\Psi\right\rangle&=&a_0\left|1,0,0,0\rangle\right\rangle+\nonumber\\
&&a_1\left|0,1,0,0\rangle\right\rangle+\nonumber\\
&&a_2\left|0,0,1,0\rangle\right\rangle+\nonumber\\
&&a_3\left|0,0,0,1\rangle\right\rangle,
\end{eqnarray}
which is under the simplest form in terms of the GHZ-criterion. If
we observe this state in a ``vertical" manner, it can be referred
to as a combination of four ``GHZ-column"s (or column for short in
this section) of which each column represents a vector consisting
of the components of the GHZ-type basis describing the system of
one group. For example, the first column of such ``GHZ-column"s is
\begin{equation}
\left(\begin{array}{c}
\left|1\rangle\right\rangle \\
\left|0\rangle\right\rangle \\
\left|0\rangle\right\rangle \\
\left|0\rangle\right\rangle
\end{array}\right).
\end{equation}
Generally, a column contains $p$ $\left|1\rangle\right\rangle$'s
and $q$ $\left|0\rangle\right\rangle$'s. In the light of the
relative ILO's, we can always make that $p\leq q$. In this case,
two columns are different iff they contain different numbers of
$\left|1\rangle\right\rangle$'s (the sequence of
$\left|1\rangle\right\rangle$'s and
$\left|0\rangle\right\rangle$'s makes no difference). Define the
direct sum of two columns by $\oplus$, e.g.,
\begin{eqnarray}
\left|\Phi\right\rangle&=& \left(\begin{array}{c}
\left|1\rangle\right\rangle \\
\left|0\rangle\right\rangle \\
\left|0\rangle\right\rangle \\
\left|0\rangle\right\rangle
\end{array}\right)\oplus
\left(\begin{array}{c}
\left|1\rangle\right\rangle \\
\left|0\rangle\right\rangle \\
\left|0\rangle\right\rangle \\
\left|0\rangle\right\rangle
\end{array}\right)=
\left(\begin{array}{c}
\left|1,0\rangle\right\rangle \\
\left|0,1\rangle\right\rangle \\
\left|0,0\rangle\right\rangle \\
\left|0,0\rangle\right\rangle
\end{array}\right)\nonumber\\
&=&a_0\left|1,0\rangle\right\rangle+a_1\left|0,1\rangle\right\rangle+
a_2\left|0,0\rangle\right\rangle+a_3\left|0,0\rangle\right\rangle.\nonumber\\
\end{eqnarray}
That is, $\oplus$ represents all ways of combinations in which two
groups will not make up one new group in the GHZ-type basis under
the GHZ-criterion. One may notice there exists another way of
combination like this,
\begin{eqnarray}
\left|\Phi\right\rangle&=&a_0\left|1,1\rangle\right\rangle+a_1\left|0,0\rangle\right\rangle
+a_2\left|0,0\rangle\right\rangle+a_3\left|0,0\rangle\right\rangle,
\end{eqnarray}
which is just the GHZ state and it does not belong to the outcome
of the operation $\oplus$. In this sense, the state (45) can be
written as
\begin{eqnarray}
\left|\Psi\right\rangle&=&
\left(\begin{array}{c}
\left|1\rangle\right\rangle \\
\left|0\rangle\right\rangle \\
\left|0\rangle\right\rangle \\
\left|0\rangle\right\rangle
\end{array}\right)\oplus
\left(\begin{array}{c}
\left|1\rangle\right\rangle \\
\left|0\rangle\right\rangle \\
\left|0\rangle\right\rangle \\
\left|0\rangle\right\rangle
\end{array}\right)\oplus
\left(\begin{array}{c}
\left|1\rangle\right\rangle \\
\left|0\rangle\right\rangle \\
\left|0\rangle\right\rangle \\
\left|0\rangle\right\rangle
\end{array}\right)\oplus
\left(\begin{array}{c}
\left|1\rangle\right\rangle \\
\left|0\rangle\right\rangle \\
\left|0\rangle\right\rangle \\
\left|0\rangle\right\rangle
\end{array}\right)\nonumber\\.
\end{eqnarray}
Notice that $\left|\Psi\right\rangle$ is composed of the same kind
of columns. In fact, the most universal form of this combination
is
\begin{equation}
\left|\Omega^{p,q}_{GHZ}\right\rangle_n\equiv\oplus\sum^{n-1}_{i=0}
(p\left|1\rangle\right\rangle,q\left|0\rangle\right\rangle)_i,1\leq
p\leq q,
\end{equation}
where
$(p\left|1\rangle\right\rangle,q\left|0\rangle\right\rangle)_i$
represents the $i$'th column containing $p$
$\left|1\rangle\right\rangle$'s and $q$
$\left|0\rangle\right\rangle$'s. By permuting the groups in the
state, identical columns are collected together and hence
$\left|\Omega_{GHZ}\right\rangle_n$ can be decomposed into the
direct sum of many
$\left|\Omega^{p^\prime,q^\prime}_{GHZ}\right\rangle_{n^\prime}$'s,
\begin{equation}
\left|\Omega_{GHZ}\right\rangle_n=\oplus\sum^{k-1}_{i=0}\left|\Omega^{p_i,q_i}_{GHZ}\right\rangle_{n_i},
p_i\leq q_i,\sum^{k-1}_{i=0}{n_i}=n.
\end{equation}
In this sense, two GHZ-type states $\left|\psi\right\rangle$ and
$\left|\phi\right\rangle$ belong to the same sort iff
$k^\psi=k^\phi,n^\psi_i=n^\phi_i,
p^\psi_i=p^\phi_i,q^\psi_i=q^\phi_i$ by corollary 2
$,i=0,1,...,k-1$, up to some permutation of the groups. In
addition, it is easy to see that two groups in two distinct
columns never make up a new group whose space is in some GHZ-type
basis. In the light of these facts, it is wise to firstly analyze
the structure of $\left|\Omega^{p,q}_{GHZ}\right\rangle_n$. We can
exemplify it as follows. Consider the case of
$|\Omega^{1,q}_{GHZ}\rangle_n$,
\begin{eqnarray}
\left(\begin{array}{c}
\left|1\rangle\right\rangle_0 \\
\left|0\rangle\right\rangle_1 \\
\left|0\rangle\right\rangle_2 \\
\cdots \\
\left|0\rangle\right\rangle_q
\end{array}\right)_0\oplus
\left(\begin{array}{c}
\left|1\rangle\right\rangle_0 \\
\left|0\rangle\right\rangle_1 \\
\left|0\rangle\right\rangle_2 \\
\cdots \\
\left|0\rangle\right\rangle_q
\end{array}\right)_1\oplus
\cdots\oplus
\left(\begin{array}{c}
\left|1\rangle\right\rangle_0 \\
\left|0\rangle\right\rangle_1 \\
\left|0\rangle\right\rangle_2 \\
\cdots \\
\left|0\rangle\right\rangle_q
\end{array}\right)_{n-1}.
\end{eqnarray}
According to the definition of direct sum $\oplus$, there is a
unique kind of combination of the same columns
$(1\left|1\rangle\right\rangle,q\left|0\rangle\right\rangle)$,
\begin{eqnarray}
|\Omega^{1,q}_{GHZ}\rangle_n&=& \left(\begin{array}{cccccc}
\left|1,0,...,0,0\rangle\right\rangle_0 \\
\left|0,1,...,0,0\rangle\right\rangle_1 \\
\cdots \\
\left|0,0,...,1,0\rangle\right\rangle_{n-2} \\
\left|0,0,...,0,1\rangle\right\rangle_{n-1} \\
\left|0,0,...,0,0\rangle\right\rangle_n \\
\cdots \\
\left|0,0,...,0,0\rangle\right\rangle_q
\end{array}\right),q\geq n-1.
\end{eqnarray}
As far as the family $|\Omega^{1,q}_{GHZ}\rangle_n$ is concerned,
there exist two inequivalent sorts of entanglement deriving from
this expression, $|\Omega^{1,n-1}_{GHZ}\rangle_n$ and
$|\Omega^{1,n}_{GHZ}\rangle_n$. However, it is generally necessary
to retain the identical terms
$\left|0,0,...,0,0\rangle\right\rangle_k,k=n,...,q$ as a result of
the direct sum. The reason lies in the decomposition of the
general GHZ-type state, e.g., consider the following state,
\begin{eqnarray}
|\Psi\rangle&=&
\left|1,0,0\rangle\right\rangle_{012}\left|0,0\rangle\right\rangle_{34}+ \nonumber\\
&&\left|0,1,0\rangle\right\rangle_{012}\left|0,0\rangle\right\rangle_{34}+ \nonumber\\
&&\left|0,0,1\rangle\right\rangle_{012}\left|1,1\rangle\right\rangle_{34}+ \nonumber\\
&&\left|0,0,0\rangle\right\rangle_{012}\left|1,1\rangle\right\rangle_{34}+ \nonumber\\
&&\left|0,0,0\rangle\right\rangle_{012}\left|0,1\rangle\right\rangle_{34}+ \nonumber\\
&&\left|0,0,0\rangle\right\rangle_{012}\left|1,0\rangle\right\rangle_{34}.
\end{eqnarray}
This state contains two kinds of columns
$(1\left|1\rangle\right\rangle,5\left|0\rangle\right\rangle)$ and
$(3\left|1\rangle\right\rangle,3\left|0\rangle\right\rangle)$, and
it is in the simplest form in terms of the GHZ-criterion. Clearly,
the ``redundant" $\left|0,0,0\rangle\right\rangle_{012}$'s is a
necessary part of the terms. This state is also a typical example
of the general GHZ-type state by the combination of different
columns.

Subsequently, let us move to the case of
$|\Omega^{2,q}_{GHZ}\rangle_n$, which is much more sophisticated
than the former one. Write out the direct sum of the columns
$(2\left|1\rangle\right\rangle,q\left|0\rangle\right\rangle)$,
\begin{eqnarray}
\left(\begin{array}{c}
\left|1\rangle\right\rangle_0 \\
\left|1\rangle\right\rangle_1 \\
\left|0\rangle\right\rangle_2 \\
\cdots \\
\left|0\rangle\right\rangle_{q+1}
\end{array}\right)_0\oplus
\left(\begin{array}{c}
\left|1\rangle\right\rangle_0 \\
\left|1\rangle\right\rangle_1 \\
\left|0\rangle\right\rangle_2 \\
\cdots \\
\left|0\rangle\right\rangle_{q+1}
\end{array}\right)_1\oplus
\cdots\oplus \left(\begin{array}{c}
\left|1\rangle\right\rangle_0 \\
\left|1\rangle\right\rangle_1 \\
\left|0\rangle\right\rangle_2 \\
\cdots \\
\left|0\rangle\right\rangle_{q+1}
\end{array}\right)_{n-1}.\nonumber\\
\end{eqnarray}
Again we need to know the possibly valued region of $q$. It is
obvious that $q_{sup}=\infty$. However, it is not easy to find out
$q_{inf}$, which is defined by that the state
$|\Omega^{2,q}_{GHZ}\rangle_n$ always holds the simplest form if
the value of $q$ is larger than it. Consider the case of $n=4$. A
possible form of the state is
\begin{eqnarray}
|\Psi_0\rangle&=&\left(\begin{array}{cccccc}
\left|1,1,1,1\rangle\right\rangle \\
\left|1,0,0,0\rangle\right\rangle \\
\left|0,1,0,0\rangle\right\rangle \\
\left|0,0,1,0\rangle\right\rangle \\
\left|0,0,0,1\rangle\right\rangle
\end{array}\right),
\end{eqnarray}
i.e., $q=3$, and it keeps the simplest form. We advance another
question, whether $|\Psi_0\rangle$ is the form in which the bases
$\left|1\rangle\right\rangle$'s distributes in the smallest
``vertical" region in all columns? Interestingly, there exists
another better arrangement,
\begin{eqnarray}
|\Psi_1\rangle&=& \left(\begin{array}{cccccc}
\left|1,1,1,0\rangle\right\rangle \\
\left|1,0,0,1\rangle\right\rangle \\
\left|0,1,0,1\rangle\right\rangle \\
\left|0,0,1,0\rangle\right\rangle \\
\left|0,0,0,0\rangle\right\rangle \\
\end{array}\right),
\end{eqnarray}
in which the bases $\left|1\rangle\right\rangle$'s only takes up 4
terms. Imagine a column to be a combination of certain
$\left|0\right\rangle$'s and $\left|1\right\rangle$'s in the
vertical region, i.e, the positions of these components make the
difference. These identical columns constructing the state then
must be in different combinations from each other. Clearly, if $q$
is less, the total number of combinations will be fewer. For the
state $|\Omega^{p,q}_{GHZ}\rangle_n$, let
\begin{equation}
\tbinom{p+q}{p}=n,
\end{equation}
whose least positive solution of $q$ is $q_{min}$. Choose the
least integer $q_{inf}\geq q_{min}$, so when $q\geq q_{inf}$ the
state can always be simplest. For the above example where
$p=2,n=4$, we have $q_{inf}=2.$ Here, the case of $q_{inf}<p$ is
also allowed, since the redundant terms
$\left|0,0,...,0\rangle\right\rangle$'s are freely added similar
to that of $|\Omega^{1,q}_{GHZ}\rangle_n$. In this sense, we
modify the definition of $|\Omega^{p,q}_{GHZ}\rangle_n$ and let
$|\omega^{p,q}_{GHZ}\rangle_n$ be the direct sum of the columns
containing $p$ components $\left|1\rangle\right\rangle$'s and
$q=q(p,n)\in[q_{inf},\infty]$ is a secondary parameter. $p+q$
indeed describes the distribution of
$\left|1\rangle\right\rangle$'s in the vertical direction. Due to
corollary 2, every value of $p+q$ leads to an essential kind of
state.

Return to the above example, where $\left|\Psi_1\right\rangle$ has
proven to be the state reaching the lowest value of $q$. One may
ask whether there exist other such states. In this expression,
there are 8 $\left|1\rangle\right\rangle$'s vertically distributed
in the manner $8=3+2+2+1$, i.e., again we meet the problem of
unordered partition, which is similar to that in the last section.
Write out other ways of partitions. For the case of $8=4+2+1+1$,
the state cannot be simplest. However, for $8=2+2+2+2$, we can
write out the second form of $|\omega^{2,2}_{GHZ}\rangle_4$,
\begin{eqnarray}
|\Psi_2\rangle&=&\left(\begin{array}{c}
\left|1,1,0,0\rangle\right\rangle \\
\left|1,0,1,0\rangle\right\rangle \\
\left|0,1,0,1\rangle\right\rangle \\
\left|0,0,1,1\rangle\right\rangle \\
\left|0,0,0,0\rangle\right\rangle
\end{array}\right),
\end{eqnarray}
and there is no the third subclass of state in the sort of
$|\omega^{2,2}_{GHZ}\rangle_4$ (the relative ILO exchanging
$\left|0\rangle\right\rangle$ and $\left|1\rangle\right\rangle$
does not generate new class). Generally, to write out all kinds of
$|\omega^{p,q}_{GHZ}\rangle_n$ where $p$ and $n$ are given and
$q\in[q_{inf},\infty]$, it suffices to analyze the terms
containing more than one $\left|1\rangle\right\rangle$, for those
containing a unique $\left|1\rangle\right\rangle$ are
automatically placed. Moreover, analyzing the number of columns
related to the terms containing more than one
$\left|1\rangle\right\rangle$ is also necessary. For example,
\begin{eqnarray}
|\omega^{2,2n-2}_{GHZ}\rangle_n&=& \left(\begin{array}{cccccc}
\left|1,0,...,0,0\rangle\right\rangle_0 \\
\left|1,0,...,0,0\rangle\right\rangle_1 \\
\left|0,1,...,0,0\rangle\right\rangle_2 \\
\left|0,1,...,0,0\rangle\right\rangle_3 \\
\cdots \\
\left|0,0,...,0,1\rangle\right\rangle_{2n-2} \\
\left|0,0,...,0,1\rangle\right\rangle_{2n-1} \\
\left|0,0,...,0,0\rangle\right\rangle \\
\cdots \\
\end{array}\right),
\end{eqnarray}
and
\begin{eqnarray}
|\omega^{2,2n-3}_{GHZ}\rangle_n&=& \left(\begin{array}{cccccc}
\left|1,1,...,0,0\rangle\right\rangle_0 \\
\left|1,0,...,0,0\rangle\right\rangle_1 \\
\left|0,1,...,0,0\rangle\right\rangle_2 \\
\cdots \\
\left|0,0,...,0,1\rangle\right\rangle_{2n-3} \\
\left|0,0,...,0,1\rangle\right\rangle_{2n-2} \\
\left|0,0,...,0,0\rangle\right\rangle \\
\cdots \\
\end{array}\right).
\end{eqnarray}
Although they are two trivial cases, the latter of which has
implies that what really functions is the first term
$\left|1,1,...,0,0\rangle\right\rangle_0$, while other terms does
not work. A complicated case from
$|\omega^{2,2n-5}_{GHZ}\rangle_n$ completely exhibits the above
statement,
\begin{eqnarray}
\left|\psi_{00}\right\rangle&=&\left(\begin{array}{c}
\overbrace{\left|1,1,1,1,0,...,0\rangle\right\rangle}^{8=4+1+1+1+1} \\
\cdots \\
\end{array}\right),\\
\left|\psi_{10}\right\rangle&=&\left(\begin{array}{c}
\overbrace{\left|1,1,1,0,0,...,0\rangle\right\rangle}^{8=3+2+1+1+1} \\
\left|0,0,0,1,1,...,0\rangle\right\rangle \\
\cdots \\
\end{array}\right),\\
\left|\psi_{11}\right\rangle&=&\left(\begin{array}{c}
\overbrace{\left|1,1,1,0,...,0\rangle\right\rangle}^{8=3+2+1+1+1} \\
\left|0,0,1,1,...,0\rangle\right\rangle \\
\cdots \\
\end{array}\right),\\
\left|\psi_{20}\right\rangle&=&\left(\begin{array}{c}
\overbrace{\left|1,1,0,0,0,0,...,0\rangle\right\rangle}^{8=2+2+2+1+1} \\
\left|0,0,1,1,0,0,...,0\rangle\right\rangle \\
\left|0,0,0,0,1,1,...,0\rangle\right\rangle \\
\cdots \\
\end{array}\right),
\end{eqnarray}
\begin{eqnarray}
\left|\psi_{21}\right\rangle&=&\left(\begin{array}{c}
\overbrace{\left|1,1,0,0,0,...,0\rangle\right\rangle}^{8=2+2+2+1+1} \\
\left|0,1,1,0,0,...,0\rangle\right\rangle \\
\left|0,0,0,1,1,...,0\rangle\right\rangle \\
\cdots \\
\end{array}\right),\\
\left|\psi_{22}\right\rangle&=&\left(\begin{array}{c}
\overbrace{\left|1,1,0,0,...,0\rangle\right\rangle}^{8=2+2+2+1+1} \\
\left|0,1,1,0,...,0\rangle\right\rangle \\
\left|0,0,1,1,...,0\rangle\right\rangle \\
\cdots \\
\end{array}\right),\\
\left|\psi_{23}\right\rangle&=&\left(\begin{array}{c}
\overbrace{\left|1,1,0,...,0\rangle\right\rangle}^{8=2+2+2+1+1} \\
\left|0,1,1,...,0\rangle\right\rangle \\
\left|1,0,1,...,0\rangle\right\rangle \\
\cdots \\
\end{array}\right).
\end{eqnarray}
The seven kinds of states are obtained firstly in terms of the
different factorizations of 8, while the number of columns plays a
key role in the second step of discrimination, e.g., the number of
related columns is 6,5,4,3 for
$\left|\psi_{20}\right\rangle,\left|\psi_{21}\right\rangle,
\left|\psi_{22}\right\rangle,\left|\psi_{23}\right\rangle$
respectively. As for the more general case, writing out all
entangled classes is indeed a complicated enumerative problem
\cite{Stanley} whose prior rule has been set up here, including
the full entanglement, the GHZ-criterion and the effective
combination of different columns, since a general GHZ-type state
$\left|\Omega_{GHZ}\right\rangle_n=\oplus\sum^{k-1}_{i=0}\left|\omega^{p_i,q_i}_{GHZ}\right\rangle_{n_i},
q_i\in[q_{inf},\infty],\sum^{k-1}_{i=0}{n_i}=n.$ Besides, the
relative ILO can make the exchange of
$\left|0\rangle\right\rangle$ and $\left|1\rangle\right\rangle$ in
the columns with the same number of $\left|0\rangle\right\rangle$
and $\left|1\rangle\right\rangle$, which is also an interesting
discipline of this problem. For all, the technique provided in the
present paper has primarily described a feasible method for this
difficult problem.

Hitherto, we have said a lot about the first step. Here we briefly
exemplify how to distill the essential sorts of entanglement from
a set of states generated by the full permutation of partitions
$N_0,N_1,...,N_{n-1}$. Consider the states $|\Psi_0\rangle$ and
$|\Psi_1\rangle$,
\begin{eqnarray}
\left(\begin{array}{cccccc}
\left|1,1,1,1\rangle\right\rangle \\
\left|1,0,0,0\rangle\right\rangle \\
\left|0,1,0,0\rangle\right\rangle \\
\left|0,0,1,0\rangle\right\rangle \\
\left|0,0,0,1\rangle\right\rangle
\end{array}\right), \mathrm{and}
\left(\begin{array}{cccccc}
\left|1,1,1,0\rangle\right\rangle \\
\left|1,0,0,1\rangle\right\rangle \\
\left|0,1,0,1\rangle\right\rangle \\
\left|0,0,1,0\rangle\right\rangle \\
\left|0,0,0,0\rangle\right\rangle
\end{array}\right).
\end{eqnarray}
Observe the position of each column. Evidently, every position of
the column in $|\Psi_0\rangle$ is equivalent and thus there still
exists only one class of entanglement for the state
$|\Psi_0\rangle$, since any state generated by some sequence of
partition leads to the same result up to a permutation of the
groups. On the other hand, the scenario differs a lot when
analyzing $|\Psi_1\rangle$, in which only the positions of column
1 and 2 are equivalent. By some simple fact of the combinatorics
we know there are at most 12 kinds of subclasses of
$\left|\Psi_1\right\rangle$, that is,
\begin{eqnarray}
(N_0,N_1,N_2,N_3),(N_0,N_1,N_3,N_2),\nonumber\\
(N_0,N_2,N_1,N_3),(N_0,N_2,N_3,N_1),\nonumber\\
(N_1,N_2,N_0,N_3),(N_1,N_2,N_3,N_0),\nonumber\\
(N_0,N_3,N_1,N_2),(N_0,N_3,N_2,N_1),\nonumber\\
(N_1,N_3,N_0,N_2),(N_1,N_3,N_2,N_0),\nonumber\\
(N_2,N_3,N_0,N_1),(N_2,N_3,N_1,N_0).
\end{eqnarray}
Practically, there may be some identical numbers in the sequence
and the repeating classes will be eliminated. Finally, by some
relative ILO's one can move away the coefficients of the state if
unnecessary.

To summarize, we have described the method of classifying the
GHZ-type state, which is actually related to the theory of
combinatorics. A significant feature of the technique is that we
can previously find out the concrete forms of relative ILO's for
this kind of entanglement by virtue of the range criterion. This
is mainly decided by the symmetry of the GHZ-type and W-type
basis, which also simplifies the calculation of entanglement of an
arbitrary pair of particles \cite{Wootters}. The relative ILO's
actually restrict the evolvement of the GHZ-type state under the
SLOCC, and there exist the analogous rule for the W-type and
GHZ-W-type state.

\section{Classifying the W-type state and the GHZ-W-type state}

This section is devoted to characterizing the W-type state by
virtue of the techniques similar to that of the last section, and
some details will be omitted if unnecessary. Subsequently, we will
primarily analyze the universal GHZ-W-type state by some examples.
Anyhow, the general classification of the GHZ-W-type state always
concerns the theory of enumerative combinatorics.

Let us start by considering the form of the relative ILO's again.
Write out $V_{0,1,...,N-1}=\bigotimes\prod^{N-1}_{i=0}V_i$ making
$R_{N0}\rightarrow R_{N0}$ on the space ${\cal H}_{P_0,P_1,\cdots
P_{N-1}}$ , where
\begin{eqnarray}
V_i=\left(\begin{array}{cc}
a_i & b_i \\
c_i & d_i
\end{array}\right)_{P_i},i=0,1,...,N-1,
\end{eqnarray}
and hence
\begin{eqnarray}
V_{0,1,...,N-1}(\alpha_0\left|0_0,0_1,...,0_{N-1}\right\rangle+\alpha_1\left|W\right\rangle_N)\nonumber\\
=\beta_0\left|0_0,0_1,...,0_{N-1}\right\rangle+\beta_1\left|W\right\rangle_N,
\alpha_0,\alpha_1,\beta_0,\beta_1\in \mathcal{C}.
\end{eqnarray}
More explicitly,
\begin{eqnarray}
\bigotimes\prod^{N-1}_{i=0}\left(\begin{array}{c}
a_i \\
c_i
\end{array}\right)_{P_i}=\beta_0\left|0_0,0_1,...,0_{N-1}\right\rangle+\beta_1\left|W\right\rangle_N,
\end{eqnarray}
and
\begin{eqnarray}
&\left(\begin{array}{c}
b_0 \\
d_0
\end{array}\right)_{P_0}
\bigotimes\prod^{N-1}_{i=1}\left(\begin{array}{c}
a_i \\
c_i
\end{array}\right)_{P_i}+
\left(\begin{array}{c}
a_0 \\
c_0
\end{array}\right)_{P_0}\bigotimes
\left(\begin{array}{c}
b_1 \\
d_1
\end{array}\right)_{P_1}&\nonumber\\
&\bigotimes\prod^{N-1}_{i=2}\left(\begin{array}{c}
a_i \\
c_i
\end{array}\right)_{P_i}
+\cdots+\bigotimes\prod^{N-2}_{i=0}\left(\begin{array}{c}
a_i \\
c_i
\end{array}\right)_{P_i}\bigotimes&
\nonumber\\
&\left(\begin{array}{c}
b_0 \\
d_0
\end{array}\right)_{P_{N-1}}
=\beta^\prime_0\left|0_0,0_1,...,0_{N-1}\right\rangle+\beta^\prime_1\left|W\right\rangle_N.&
\end{eqnarray}
By the first expression, $c_ic_j=0,i,j=0,1,...,N-1,i\neq j$, so
every $c_i=0$. Without loss of generality, let
$a_i=1,i=0,1,...,N-1.$ Then by the second expression,
$d_i=d_j,i,j=0,1,...,N-1$. In other words, the relative
ILO can be expressed as \\
$V_{0,1,...,N-1}=\bigotimes\prod^{N-1}_{i=0}\left(\begin{array}{cc}
1 & b_i \\
0 & x
\end{array}\right)_{P_i},x\neq0$,\\
so the total effect on the system is
\begin{eqnarray}
V_{0,1,...,N-1}\left|0\rangle\right\rangle_N&=&\left|0\rangle\right\rangle_N,\\
V_{0,1,...,N-1}\left|W\right\rangle_N&=&x\left|W\right\rangle_N+(\sum^{N-1}_{i=0}b_i)\left|0\rangle\right\rangle_N.
\end{eqnarray}
Therefore for the W-type state, the essential change by the
relative ILO is
$\left|W\right\rangle\rightarrow\left|W\right\rangle+\alpha\left|0\rangle\right\rangle$,
where $\alpha$ is an arbitrarily determined parameter in advance.
Besides, the base $\left|0\rangle\right\rangle$ remains invariant,
and the coefficients in the state can be efficiently reduced by
the universal factor $x$ in the expression of $V_{0,1,...,N-1}$.
Regard each product form
$\bigotimes\prod^{k-1}_{i=0}\left|0\rangle\right\rangle_{N_i}
\bigotimes\prod^{n-1}_{i=k}\left|W\right\rangle_{N_i}$ as a term
in the state $\left|\Omega_W\right\rangle_n$. then
$V_{0,1,...,N-1}$ does not change the $number$ of the terms
containing the most $\left|W\right\rangle$'s. That is,

\textit{Corollary 3}. Suppose two W-type states are equivalent
under SLOCC, then they have the same number of the terms
containing the most $\left|W\right\rangle$'s. We call it the
highest-term. \hspace*{\fill}$\blacksquare$

For instance, consider the two-group W-type state,
\begin{eqnarray}
\left|\Omega_W\right\rangle_2=\left|W\right\rangle_{N_0}\left|W\right\rangle_{N_1}+
a_1\left|W\right\rangle_{N_0}\left|0\rangle\right\rangle_{N_1}\nonumber\\
+a_2\left|0\rangle\right\rangle_{N_0}\left|W\right\rangle_{N_1}+
a_3\left|0\rangle\right\rangle_{N_0}\left|0\rangle\right\rangle_{N_1}.
\end{eqnarray}
By using of two ILO's making
$\left|W\right\rangle_{N_0}\rightarrow\left|W\right\rangle_{N_0}-a_2\left|0\rangle\right\rangle_{N_0}$
and
$\left|W\right\rangle_{N_1}\rightarrow\left|W\right\rangle_{N_0}-a_1\left|0\rangle\right\rangle_{N_1}$
respectively, we obtain the unique class of the two-group W-type
state
\begin{equation}
\left|\Omega_W\right\rangle_2=\left|W\right\rangle_{N_0}\left|W\right\rangle_{N_1}+
\left|0\rangle\right\rangle_{N_0}\left|0\rangle\right\rangle_{N_1},
\end{equation}
where the coefficient has been moved away by some ILO $O^{N_0}_1$.
Evidently, the highest-term
$\left|W\right\rangle_{N_0}\left|W\right\rangle_{N_1}$ does not
disappear under the ILO's. Moreover, the W-type state has no
decomposition form similar to that of the GHZ-type state.
Fortunately, the procedure of classification thereof still serves.
Let
$\bigotimes\prod^{n-1}_{i=0}\left|x_i\rangle\right\rangle_{N_i}=
\left|x_0,x_1,...,x_{n-1}\rangle\right\rangle,x_i=0,W$, to firstly
find out all sorts of entanglement under an average division
$N_i=N_j,i,j=0,1,...,n-1$, and then get all states by a full
permutation of the partitions $N_0,N_1,...,N_{n-1}$. After
eliminating the repeating cases and moving away the coefficients
as many as possible, the essential classes of states are obtained.
Here, we only describes the first step and omit the trivial parts,
which one can deal with by following the techniques in the last
section. Write out the general form of the W-type state,
\begin{eqnarray}
\left|\Omega_W\right\rangle_n=a_0\left|W,W,...,W\rangle\right\rangle+
\sum^{\tbinom{n}{1}}_{i=1}a_{1,i}P_i(\left|W,W,...,W,0\rangle\right\rangle)\nonumber\\
+\sum^{\tbinom{n}{2}}_{i=1}a_{2,i}P_i(\left|W,W,...,W,0,0\rangle\right\rangle)+\cdots+\nonumber\\
\sum^{\tbinom{n}{n-1}}_{i=1}a_{n-1,i}P_i(\left|W,0,...,0,0\rangle\right\rangle)
+a_n\left|0,0,...,0,0\rangle\right\rangle,\nonumber\\
\end{eqnarray}
where $\{P_i\}$ is the set of all $\tbinom{N}{m}$ distinct
permutations of the groups. Due to corollary 3, we can catalog
this state by the changing number of highest-terms. Define
$\left|\Omega^{p,q}_W\right\rangle_n$ as the W-type state
containing $p$ highest-terms
$P_i(|\overbrace{W,...,W}^q,0,...,0\rangle\rangle), 1\leq
p\leq\tbinom{n}{q},2\leq q\leq n$, e.g., there is one term
$\left|W,W,...,W\rangle\right\rangle$, $n$ terms
$P_i(\left|W,W,...,W,0\rangle\right\rangle)$, etc. Totally, there
exist
$M=1+\tbinom{n}{1}+\tbinom{n}{2}+\cdots+\tbinom{n}{n-2}=2^n-n-1$
main classes of W-type entanglement. However, even for a class
with a certain number of highest-terms, there may still exist some
subclasses of states by permutation similar to that in last
section. Observe the following two states on which some ILO's has
been operated,
\begin{eqnarray}
|\Omega^{2,2}_W\rangle^A_4&=&\left|W,W,0,0\rangle\right\rangle+
\left|0,0,W,W\rangle\right\rangle+\alpha\left|0,0,0,0\rangle\right\rangle,\nonumber\\
|\Omega^{2,2}_W\rangle^B_4&=&\left|W,W,0,0\rangle\right\rangle+\left|W,0,W,0\rangle\right\rangle+\nonumber\\
&&\left|0,0,0,W\rangle\right\rangle+\left|0,0,W,0\rangle\right\rangle.
\end{eqnarray}
Notice the parameter $\alpha=0$, or $1$ is necessary here, since
each of them leads to an essential sort of W-type state, which can
be easily checked by the relative ILO's. The main point we
emphasize is the inequivalence of $|\Omega^{2,2}_W\rangle^A_4$ and
$|\Omega^{2,2}_W\rangle^B_4$. In the light of the W-criterion,
both of them are in the simplest form. Similar to the case of the
GHZ-type state, the components $\left|W\right\rangle$'s distribute
in all four columns in $|\Omega^{2,2}_W\rangle^A_4$, while in only
three columns in $|\Omega^{2,2}_W\rangle^B_4$, so they are
inequivalent under SLOCC. This distinction derives from the
different arrangement of the components $\left|W\right\rangle$'s.
However, the intrinsic change happens in the lower-terms, i.e.,
those containing less $\left|W\right\rangle$'s. As the relative
ILO's can make $\forall
\alpha,\left|W\right\rangle\rightarrow\left|W\right\rangle+\alpha\left|0\rangle\right\rangle$,
the number of the lower-terms generally can not be the evidence of
inequivalent states, and finding out a regular arrangement of
$\left|W\right\rangle$'s remains a difficult problem. We analyze
some situations to illustrate it.

(i) The state $|\Omega^{1,n}_W\rangle_n$. By performing the ILO's
$\left|W\right\rangle_{N_i}\rightarrow\left|W\right\rangle_{N_i}
+\alpha_i\left|0\rangle\right\rangle_{N_i},i=0,1,...,n-1$, it is
in a more succinct form,
\begin{eqnarray}
\left|\Omega^{1,n}_W\right\rangle_n=\left|W,W,...,W\rangle\right\rangle
+\sum^{\tbinom{n}{2}}_{i=1}a_{2,i}P_i(\left|W,W,...,W,0,0\rangle\right\rangle)\nonumber\\
+\cdots+\sum^{\tbinom{n}{n-1}}_{i=1}a_{n-1,i}P_i(\left|W,0,...,0,0\rangle\right\rangle)
+a_n\left|0,0,...,0,0\rangle\right\rangle.\nonumber\\
\end{eqnarray}
The above state has an important character, i.e., the relative
ILO's on this state must make
$\left|W\right\rangle\rightarrow\left|W\right\rangle$ and
$\left|0\rangle\right\rangle\rightarrow\left|0\rangle\right\rangle$,
otherwise some term $P_i(\left|W,W,...,W,W,0\rangle\right\rangle)$
will always appear. In this case, once more the different numbers
of the lower-terms lead to inequivalent classes of W-type states,
which implies the existing techniques can be applied to this
scenario.

(ii) The state $|\Omega^{p,n-1}_W\rangle_n$. The part of the
highest-terms can be written in terms of columns,
\begin{eqnarray}
\left(\begin{array}{cccccc}
\left|W,W,W,\cdots,W,W,0\rangle\right\rangle_0 \\
\left|W,W,W,\cdots,W,0,W\rangle\right\rangle_1 \\
\left|W,W,W,\cdots,0,W,W\rangle\right\rangle_2 \\
\cdots \\
|W,\cdots,W,0,\overbrace{W,\cdots,W}^{p-1}\rangle\rangle_{p-1} \\
\end{array}\right),
\end{eqnarray}
which is the unique form of the state
$|\Omega^{p,n-1}_W\rangle_n$. Interestingly, the backward main
diagonal of the right $p$ columns is composed of identical
$\left|0\rangle\right\rangle$'s. Besides, only the state
$|\Omega^{2,n-1}_W\rangle_n$ could be not in the simplest form by
the W-criterion. However, it is difficult to decide the form of
the relative ILO's, for any ILO making
$\left|W\right\rangle\rightarrow\left|W\right\rangle+\alpha\left|0\rangle\right\rangle$
will affect all the lower-terms, while the unknown coefficients
are a lot. This awful fact appears in all other cases of
$|\Omega^{p,q}_W\rangle_n$. Nonetheless, we still can deal with
many classes by the existing techniques, e.g.,
$|\Omega^{2,2}_W\rangle^A_4$ and $|\Omega^{2,2}_W\rangle^B_4$ are
the only two types of states of $|\Omega^{2,2}_W\rangle_4$. A more
involved case is the state $|\Omega^{3,2}_W\rangle_4$. We
demonstrate the three distinct forms of the highest terms,
\begin{eqnarray}
\left(\begin{array}{cc}
\overbrace{|W,W,0,0}^{6=3+1+1+1}\rangle\rangle_0 \\
\left|W,0,W,0\rangle\right\rangle_1 \\
\left|W,0,0,W\rangle\right\rangle_2 \\
\end{array}\right),
\left(\begin{array}{cc}
\overbrace{|W,W,0,0}^{6=2+2+1+1}\rangle\rangle_0 \\
\left|W,0,W,0\rangle\right\rangle_1 \\
\left|0,W,0,W\rangle\right\rangle_2 \\
\end{array}\right),
\end{eqnarray}
and
\begin{eqnarray}
\left(\begin{array}{cc}
\overbrace{|W,W,0,0}^{6=2+2+2}\rangle\rangle_0 \\
\left|0,W,W,0\rangle\right\rangle_1 \\
\left|W,0,W,0\rangle\right\rangle_2 \\
\end{array}\right).
\end{eqnarray}
By some simple ILO's, the three subclasses of
$|\Omega^{3,2}_W\rangle_4$ can be briefly written as
\begin{eqnarray}
|\Omega^{3,2}_W\rangle^A_4=|W,W,0,0\rangle\rangle+|W,0,W,0\rangle\rangle+|W,0,0,W\rangle\rangle+\nonumber\\
a_0|0,0,W,0\rangle\rangle+a_1|0,0,0,W\rangle\rangle+a_2|0,0,0,0\rangle\rangle,\nonumber\\
|\Omega^{3,2}_W\rangle^B_4=|W,W,0,0\rangle\rangle+|W,0,W,0\rangle\rangle+|0,W,0,W\rangle\rangle+\nonumber\\
b_0|0,0,W,0\rangle\rangle+b_1|0,0,0,W\rangle\rangle+b_2|0,0,0,0\rangle\rangle,\nonumber\\
|\Omega^{3,2}_W\rangle^C_4=|W,W,0,0\rangle\rangle+|0,W,W,0\rangle\rangle+|W,0,W,0\rangle\rangle+\nonumber\\
c_0|0,0,W,0\rangle\rangle+|0,0,0,W\rangle\rangle+c_2|0,0,0,0\rangle\rangle,\nonumber\\
\end{eqnarray}
in each of them the coefficients characterize the structure of
this state. Moreover, we can simplify the third expression by the
ILO's
$\left|W\right\rangle_{N_i}\rightarrow\left|W\right\rangle_{N_i}
+x_i\left|0\rangle\right\rangle_{N_i},i=0,1,2,3$ such that
\begin{eqnarray}
|\Omega^{3,2}_W\rangle^C_4=|W,W,0,0\rangle\rangle+|0,W,W,0\rangle\rangle+|W,0,W,0\rangle\rangle+\nonumber\\
+(x_1+x_2)|W,0,0,0\rangle\rangle+(x_0+x_2)|0,W,0,0\rangle\rangle+\nonumber\\
(x_0+x_1+c_0)|0,0,W,0\rangle\rangle+|0,0,0,W\rangle\rangle,\nonumber\\
\end{eqnarray}
and supposing $x_1+x_2=x_0+x_2=x_0+x_1+c_0=0$ leads to
\begin{eqnarray}
|\Omega^{3,2}_W\rangle^C_4&=&|W,W,0,0\rangle\rangle+|0,W,W,0\rangle\rangle+\nonumber\\
&&|W,0,W,0\rangle\rangle+|0,0,0,W\rangle\rangle,
\end{eqnarray}
which is the most succinct form of $|\Omega^{3,2}_W\rangle^C_4$.
One can simplify other expressions above in the similar way,
despite some tedious algebra.

Of all above, we have analyzed two typical families of the
GHZ-W-type state, the GHZ-type state and the W-type state.
Generally, the state $\left|\Omega_{GHZ-W}\right\rangle_{k,n-k}$
can be obtained by the direct sum of the state
$\left|\Omega_{GHZ}\right\rangle_k$ and
$\left|\Omega_W\right\rangle_{n-k}$,
\begin{equation}
\left|\Omega_{GHZ-W}\right\rangle_{k,n-k}=\left|\Omega_{GHZ}\right\rangle_k\oplus\left|\Omega_W\right\rangle_{n-k}.
\end{equation}
For convenience, we can regard the two states as two universal
kinds of ``basis", i.e., the universal GHZ-type and universal
W-type basis respectively. In this sense, the operation $\oplus$
means all matches of every pairwise components in the two kinds of
basis respectively. Besides, the relative ILO's for the two parts
are known and either of the distinct classes of the two basis will
lead to the essential classes of the GHZ-W-type state.
Furthermore, the rules for determining whether the state is
simplest still serve here. Due to the operation of the direct sum,
the part of the W-type basis becomes more unrestricted. For
instance, the state $|\Omega^{p,1}_W\rangle_n$ always cannot be in
the simplest form since
\begin{eqnarray}
\left(\begin{array}{cccccc}
\left|0,0,0,\cdots,0,0,W\rangle\right\rangle_0 \\
\left|0,0,0,\cdots,0,W,0\rangle\right\rangle_1 \\
\left|0,0,0,\cdots,W,0,0\rangle\right\rangle_2 \\
\cdots \\
|0,\cdots,0,W,\overbrace{0,\cdots,0}^{p-1}\rangle\rangle_{p-1} \\
\left|0,0,0,\cdots,0,0,0\rangle\right\rangle_p \\
\cdots \\
\end{array}\right)\sim
\left|0\rangle\right\rangle_{\sum\limits^{n-p-1}_{i=0}N_i}
\left|W\right\rangle_{\sum\limits^{n-1}_{i=n-p}N_i}.\nonumber\\
\end{eqnarray}
However, the following state remains simplest,
\begin{eqnarray}
\left|\Psi\right\rangle&=&|\omega^{1,2}_{GHZ}\rangle_3\oplus|\Omega^{2,1}_W\rangle_2\nonumber\\
&=&|1,0,0\rangle\rangle(a_0|W,0\rangle\rangle+a_1|0,W\rangle\rangle+a_2|0,0\rangle\rangle)+\nonumber\\
&&|0,1,0\rangle\rangle(b_0|W,0\rangle\rangle+b_1|0,W\rangle\rangle+b_2|0,0\rangle\rangle)+\nonumber\\
&&|0,0,1\rangle\rangle(c_0|W,0\rangle\rangle+c_1|0,W\rangle\rangle+c_2|0,0\rangle\rangle)+\nonumber\\
&&|0,0,0\rangle\rangle(d_0|W,0\rangle\rangle+d_1|0,W\rangle\rangle+d_2|0,0\rangle\rangle),
\end{eqnarray}
when the coefficients are appropriately given, such that it
satisfies the W-criterion. We are interested in the classification
of this universal type of multiqubit state. Since the position of
the columns in the part of the GHZ-basis is identical, there
exists some repeating cases when setting the coefficients. By
corollary 3, the number of the highest terms does not change under
the relative ILO's. In this case, the numbers of
$P_i(|W,0\rangle\rangle)$'s in each bracket are fixed. Denote
these numbers in the prior three brackets by $(p_0,p_1,p_2)$, and
the number ``0" means the highest term is $|0,0\rangle\rangle$.
Then there are nine main classes of the state
$\left|\Psi\right\rangle$, whose numbers are
\begin{eqnarray}
(2,2,2),(2,2,1),(2,2,0),(2,1,1),(2,1,0),\nonumber\\
(2,0,0),(1,1,1),(1,1,0),(1,0,0),
\end{eqnarray}
up to some permutation of the groups (the combination $(0,0,0)$
does not lead to the simplest form). More explicitly, there are
two subclasses in the states with numbers $(2,1,1)$, $(1,1,1)$,
and $(1,1,0)$. By some simple ILO's we can write out the succinct
forms as follows,
\begin{eqnarray}
\left|\Psi\right\rangle^A_{(2,1,1)}=
|1,0,0\rangle\rangle(|W,0\rangle\rangle+|0,W\rangle\rangle)+\nonumber\\
|0,1,0\rangle\rangle(|W,0\rangle\rangle+b_2|0,0\rangle\rangle)+|0,0,1\rangle\rangle|W,0\rangle\rangle\nonumber\\
+|0,0,0\rangle\rangle(d_0|W,0\rangle\rangle+d_1|0,W\rangle\rangle+d_2|0,0\rangle\rangle),\\
\left|\Psi\right\rangle^B_{(2,1,1)}=
|1,0,0\rangle\rangle(|W,0\rangle\rangle+|0,W\rangle\rangle)+\nonumber\\
|0,1,0\rangle\rangle(|W,0\rangle\rangle+b_2|0,0\rangle\rangle)+|0,0,1\rangle\rangle|0,W\rangle\rangle\nonumber\\
+|0,0,0\rangle\rangle(d_0|W,0\rangle\rangle+d_1|0,W\rangle\rangle+d_2|0,0\rangle\rangle),\\
\left|\Psi\right\rangle^A_{(1,1,1)}=
|1,0,0\rangle\rangle(|W,0\rangle\rangle+a_2|0,0\rangle\rangle)+\nonumber\\
|0,1,0\rangle\rangle(|W,0\rangle\rangle+b_2|0,0\rangle\rangle)+|0,0,1\rangle\rangle|W,0\rangle\rangle\nonumber\\
+|0,0,0\rangle\rangle(d_0|W,0\rangle\rangle+|0,W\rangle\rangle),\\
\left|\Psi\right\rangle^B_{(1,1,1)}=
|1,0,0\rangle\rangle(|W,0\rangle\rangle+a_2|0,0\rangle\rangle)+\nonumber\\
|0,1,0\rangle\rangle|W,0\rangle\rangle+|0,0,1\rangle\rangle|0,W\rangle\rangle+\nonumber\\
|0,0,0\rangle\rangle(d_0|W,0\rangle\rangle+d_1|0,W\rangle\rangle+d_2|0,0\rangle\rangle),
\end{eqnarray}
\begin{eqnarray}
\left|\Psi\right\rangle^A_{(1,1,0)}&=&
|1,0,0\rangle\rangle(|W,0\rangle\rangle+a_2|0,0\rangle\rangle)\nonumber\\
&+&|0,1,0\rangle\rangle|W,0\rangle\rangle+|0,0,1\rangle\rangle|0,0\rangle\rangle\nonumber\\
&+&|0,0,0\rangle\rangle(d_0|W,0\rangle\rangle+|0,W\rangle\rangle),\\
\left|\Psi\right\rangle^B_{(1,1,0)}&=&
|1,0,0\rangle\rangle|W,0\rangle\rangle+|0,1,0\rangle\rangle|0,W\rangle\rangle\nonumber\\
&+&|0,0,1\rangle\rangle|0,0\rangle\rangle+|0,0,0\rangle\rangle\nonumber\\
&(&d_0|W,0\rangle\rangle+d_1|0,W\rangle\rangle+d_2|0,0\rangle\rangle).
\end{eqnarray}
One can also list other states containing only one subclasses by
the similar techniques. Predictably, the general GHZ-W-type state
can be both qualitatively and quantitatively characterized by the
techniques developed by us.

\section{conclusion}

In this paper, first we introduced the concept of the
SLOCC-equivalent basis (SEB), and proposed two general SEBs, the
GHZ-type and the W-type basis in the multiqubit space. By virtue
of the two SEBs, we proved that the GHZ state and the W state are
the only two types consisting of the basis $R_{20}$ and $R_{21}$
in any two-qubit subspace respectively. Subsequently, a universal
type of pure multiqubit state consisting of the GHZ-type and
W-type basis, i.e., the GHZ-W-type state was carefully
investigated. We proposed the condition on which this state can be
fully entangled and be in the simplest form. The main purpose of
this paper is to classify the GHZ-W-type multiqubit state, which
can be realized by characterizing the GHZ-type and W-type state
respectively. Our result has shown that the full understanding of
this type of multiqubit state tightly relates to the theory of
combinatorics.

There are several aspects for the future work. First, more
algebraic effort is required, especially for the case of the
W-type state. The classification of the GHZ-W-type state proves to
be a difficult problem in the enumerative combinatorics, and it is
expected that a more explicit method can be given. Although we
have classified the entanglement under the SLOCC criterion, it is
easy to modify our argument so that it becomes the LOCC criterion,
since most of the relative ILO's are diagonal or anti-diagonal
(one can set the entries $b_i=0,\forall i$ in the ILO's for the
W-type state). Therefore, the classification becomes utterly
precise and most of our conclusions still serve. Second, recently
a few types of multiqubit entanglement have been demonstrated by
the spontaneous parametric down-conversion (SPDC), such as the
4-qubit GHZ state and 5-qubit GHZ state \cite{Pan}. In particular,
Weinfurther \textit{et al.} \cite{Weinfurter} has realized a
superposition of a four photon GHZ state and a product of two EPR
pairs by using the type-II down-conversion \cite{Kwiat},
\begin{equation}
|\Psi^{(4)}\rangle=\sqrt{\frac23}|\mathrm{GHZ}\rangle_{aa^{\prime}bb^{\prime}}
-\sqrt{\frac13}|\mathrm{EPR}\rangle_{aa^{\prime}}|\mathrm{EPR}\rangle_{bb^{\prime}}.
\end{equation}
Here, the $\mathrm{EPR}$ state is the Bell state
$(1/\sqrt2)(\left|01\right\rangle_{xx^{\prime}}-\left|10\right\rangle_{xx^{\prime}})$
with $x=a,b.$ Evidently, this state has a more complicated
configuration than the GHZ-W-type state since it consists of the
rank-3 basis in each bipartite subspace. It is thus possible to
realize the 4-qubit GHZ-W-type state by the similar experimental
setup, for it can be seen as the superposition of a product of two
EPR pairs and another replaceable term, which can be
$\left|0000\right\rangle$,
$\left|0000\right\rangle+\left|0011\right\rangle$, etc. An
alternative method of producing it can be obtained by the joint
measurement of the state of the composite system, e.g., the state
$\left|\Psi\right\rangle_{sub}$ in \cite{Deng2} is indeed the
4-qubit GHZ-type state. Finally, the character of the GHZ-W-type
state is also an interesting topic, such as the multiqubit
entanglement measures \cite{Wong} and the robustness of it
\cite{Tarrach}.

The work was partly supported by the NNSF of China Grant
No.90503009 and 973 Program Grant No.2005CB724508.


\begin{references}
\bibitem{notation1}
In general, it is hard to seek the generic form of all rank-2 SEBs
in the multiqubit space, because too many parameters appear as the
number of parties increases. We will apply the theory of entangled
bases to the characterization of 5-qubit entanglement in a
posterior work.
\end{references}
\end{document}